\documentclass[a4paper,11pt]{article}
\usepackage{fullpage}
\usepackage{times}
\usepackage{subfigure}
\usepackage{vmargin}
\usepackage{fancyheadings}
\usepackage[cp1251]{inputenc}
\usepackage[english]{babel}
\usepackage{graphicx}
\usepackage{color}
\usepackage{cite}
\usepackage{multicol}
\usepackage{amsmath,amssymb,amsthm,enumerate,bbm,color,verbatim,setspace}
\usepackage{color,colortbl}
\definecolor{darkishgreen}{RGB}{39,203,22}
\definecolor{RoyalBlue}{rgb}{0.88,1,1}
\definecolor{Gray}{gray}{0.99}
\definecolor{Orange}{RGB}{139, 0, 255}
\definecolor{Yellow}{rgb}{1, 1, 0}
\definecolor{strongRed}{RGB}{230,6,6}

\date{}
\title{Statistical Properties of European Languages and Voynich Manuscript Analysis}
\author{Andronik Arutyunov$^{1}$, Leonid Borisov$^{2}$, Sergey Fedorov$^{2}$,\\
 Anastasiya Ivchenko$^{2}$, Elizabeth Kirina-Lilinskaya$^{2}$, Yurii Orlov$^{2*}$,\\
 Konstantin  Osminin$^{2}$,  Sergey Shilin$^{3}$, Dmitriy Zeniuk$^{2}$\\
\\
\\
$^1$ Peoples' Friendship University of Russia, Moscow, Russia\\
$^2$ Keldysh Institute of Applied Mathematics of RAS, Moscow, Russia\\
$^3$ Moscow Institute of Physics and Technology, Moscow, Russia\\
\\
$^2{^{*}}$Corresponding e-mail: yuno@kiam.ru\\}
\date{}
\begin{document}
\newcolumntype{g}{>{\columncolor{Gray}}c}
\maketitle
\pagenumbering{arabic}
\begin{abstract}\noindent
The statistical properties of letters frequencies in European literature texts are investigated. The determination of logarithmic dependence of letters sequence for one-language and two-language texts are examined. The pare of languages is suggested for Voynich Manuscript. The internal structure of Manuscript is considered. The spectral portraits of two-letters distribution are constructed.
\end{abstract}

\medskip

\newpage
\tableofcontents
\newpage
\section{Introduction}
\par The main aim of the present paper is a statistic analysis of literary texts written in European languages  (we will consider Indo-European and Uralic language families). We construct a letter distribution using large texts (more than 100,000 characters)  in order to identify similar properties in the relevant lexicons.
\par This paper is partly based on results obtained in the preprint \cite{ORIG} and methods from the monograph \cite{OrOs} where it could be found some results of letter frequency anylisis in European languages.  In particular, it was found that for the most languages the dependence of   ordered frequencies is accuratly logarithmic, its determination is more than 0.98. Parameters of the logarithmic dependence are determined by the number of characters in an alphabet and allow the interpretation of the redundancy or the failure.
\par Our interest is in a deviation from logarithmic dependence in letter statistics for bilingual texts (written in a full alphabet$/$ consonants only $/$ written in a constructed language). Would it be correct using statistics of the frequency of letters to make sufficiently reliable supposition about texts language? This question has arisen from paper authors' Voynich manuscript discussion.
\par The Voynich manuscript (VM \cite{Shailor}) -- is a hand-written codex, dating from the XVI c. It consists of over 170,000 characters referred to as letters, which are  united by transcriptioners in 22 distinct characters. These characters  are not elements of any known alphabet. At the present time the manuscript is kept in the Beinecke Library and has the status of a cryptographic puzzle. 
\par  Numerous studies in order to decrypt the text carried out more than a hundred years and are still unsuccessfully. Versions of the authorship, content and language of the manuscript (you can found a review in [4-6]) in our opinion are not supported enough by the full-fledged statistical studies.
\par We have to emphasize that our aim is not to decode the manuscript. We do not analyze the vocabulary so we do not discuss the semantic component of the text.  We will try to get the answer to the following question: Is the VM a encrypted  meaningful text (and so in what language is it written?) or is a hoax, i. e. meaningless set of characters? It may seem that the answer is required a decryption of the text but in reality this is not necessary. Firstly it  is important to find out if there is a common statistical properties in texts without knowledge of which it is impossible to make the necessary simulation. Our studies show that such properties exist.
\par There is no consensus how many characters VM is consists of. We will consider two transcriptions of VM: ``European transcription" (EVA \cite{LZ})  and transcription Takahashi \cite{Tak}. Both of them translate manuscript symbols into Latin alphabet, but  with different frequencies of symbols. (the fact is we can not  interpret  many of caracters in MV uniquely). An any case we will not discuss a correctness of these transcriptions. Our goal is to study statistic properties of them. Therefore it should be remembered by speaking about the ``MV structure" or ``language of its parts" we will intend exactly these transcriptions.
\par Researchers have proposed numerous hypotheses about the structure of the Manuscript. There are some known theories:
\begin{itemize}
\item it was written with permutation of letters 
\item two letters of the well-known alphabet correspond to one character of the manuscript;
\item There is a key without which you can not read the text because the same characters in different parts of the manuscript correspond to different letters
\item the manuscript is an encoded two-language  text;
\item vowels have been removed from the originally meaningful text ;
\item the text contains false spaces between words.
\end{itemize}

\par At the same time in various  concepts (unproven in the statistical sense) for the role of the original  language are proposed: Hebrew, Spanish, Russian, Manchurian, Vietnamese and much more (even Arabic or ``something Indian''). At the same time we can consider the existence of false spaces as a real component of Manuscript structure. In this instance a decryption may be very problematic. 
In addition, it should be noted that if the text contains no vowels, the vowel recovery is not uniquely.

\par One further remark is that pages of Manuscript could be numbered not by the author and the oder of ``words''  can be changed. There are also no any evidences that VM is a one text,  not two or three texts under one cover -- throughout the text a handwriting is not the same. 
\par One must refrain from the idea of seeing words in VM.  We can consider only character statistics assuming that every character is a letter.  
\par Studies \cite{OrOs} have shown that symbol distribution of frequency of occurrence is  a strong characteristic of language not an author or a text theme. 
\par It is assumed that the distribution of text mixtures in two languages will be equally stable. It will be judged by the level of its determination with some model distribution the equity participation of different languages in these bilingual texts
\par It is possible that the text character distributions are stable within the same language group  and for  texts written in mixed language from two different language groups (e.g. one part is written in English, another in French), distributions will be unstable. In this case we can talk about natural clusterization of kindred languages for texts written in arbitrary two-language mixture. This Clusterization is based on the principle of the text closeness by frequency-ordered characters distribution. It is also very interesting to compare texts written in languages from different language groups but  in the one alphabetic system, for example, in Hungarian (Uralic family, Finno-Ugric branch) and English (Indo-European family, the West Germanic branch). In addition we will consider some constructed languages in oder to  consider the case when  VM is written in a constructed laguage.
\par In this way, our paper deals with the analysis of those assumptions and study invariant properties of European languages.  For finding linguistic invariants the following statistics are used:  the distance between distributions of odered empirical frequencies of letter combinations in norm $L_1$;  determination level of logarithmic approximations of one-letter distributions for texts without vocalisation; Hurst index distribution for a series of the number of letters concluded between the two most frequently encountered same letters;  spectral matrix portrait of two-letter combinations.  These indicators have allowed to make the formal clusterization of languages from Indo-European family.  As result our clusters have coincided with groups formed on the basis of studies in Historical Linguistics.

{\footnotesize
\begin{table}
\caption{{\footnotesize MV frequencies}}
\begin{center}
\begin{tabular}{|g|g|g|}
\hline
Symbol & EVA & Takahashi\\
\hline
a & 0.07456 & 0.07641\\
\hline
b & & 0.00051\\
\hline
c & 0.06951 & 0.00254\\
\hline
d & 0.06773 & 0.00269\\
\hline
e & 0.10478 & 0.12940\\
\hline
f & 0.00264 & 0.00598\\
\hline
g & 0.00050 & 0.00019\\
\hline
h & 0.09322 & 0.11559\\
\hline
i & 0.06125 & 0.01472\\
\hline
k & 0.05708 & 0.02901\\
\hline
l & 0.05491 & 0.05624\\
\hline
m & 0.00583 & 0.03713\\
\hline
n & 0.03206 & 0.03988\\
\hline
o & 0.13296 & 0.13616\\
\hline
p & 0.00851 & \\
\hline
q & 0.02831 & 0.00870\\
\hline
r & 0.03893 & 0.02402 \\
\hline
s & 0.03857 & 0.01541\\
\hline
t & 0.03625 & 0.12789\\
\hline
u & & 0.00011\\
\hline
v & 0.00005& \\
\hline
w & &0.08296\\
\hline
x & 0.00018 & \\
\hline
y & 0.09217 & 0.09445\\
\hline
z & 0.00001 & 0.00001\\
\hline
\end{tabular}
\end{center}
\end{table}
}
\section {Manuscript transcriptions statistics}
\par There are frequencies for MV symbols in Table 1  which are obtained as a transcription of  manuscript  characters into the Latin alphabet. There are two versions of transcription: EVA and Takahashi. In calculating we did not  distinguish uppercase and lowercase letters and did not treat  empty spaces as  special symbols.
\par In our further analysis we will focus on the construction of the VM symbol distribution of frequency of occurrence. We will compare it with analogical distributions in European languages and will revial the deviation from the  level of determination of approximated dependence. Then we will determine the distance between actual distribution and its approximation  in norm $L_1$.
\par  There are symbol descending ordered frequencies in Fig.1. These distributions are closed in sens of determination level of approximated dependence (0.93), but they are different substantially in details . According to the studies [2],  the EVA graph (red line) is typical for Germanic group, or rather West Germanic languages. Takahashi graph (green line)  is typical for Slavic and Romance languages (Fig. 2) and also for North Germanic languages. The distance between these transcriptions (with descending ordered frequencies) in norm  $L_1$ is equal to 0.26 that is in 3 times more than between distributions of texts without vocalisation in one language family, in 10 times more between texts with full alphabet. It means that each of such transcriptions corresponds with fundamentally different analysis of VM, so there is no possibility for use both translations in oder to statistics elboration.

\begin{figure}[ht]
\begin{center}
\begin{minipage}[ht]{0.70\linewidth}
\includegraphics[width=1\linewidth]{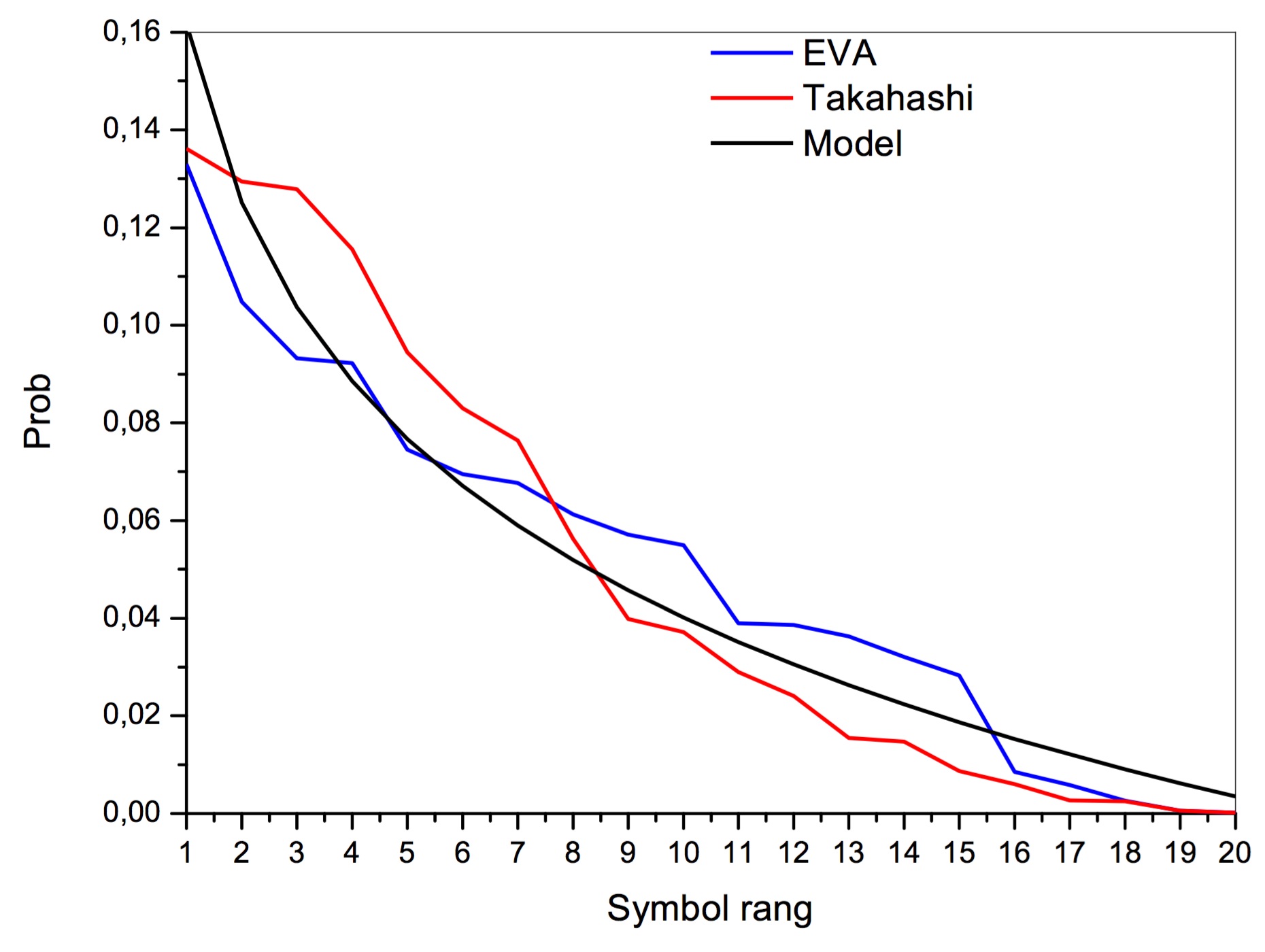}
\vspace{-4mm}
\caption{{\footnotesize Ordered frequency of two VM-transcriptions and logarithmic model approximation}}
\label{fig_1}
\end{minipage}
\end{center}
\end{figure}
\par For most of modern languages in Indo-European family logarithmic dependence  of letter frequency on its rang is typical with accuracy more than 0.98. The level of determination in texts without vocalisation is slightly lower -- 0.96 (Fig 2-5, Table 2).
Actual distribution of odered frequencies for texts written in the same language differs from logarithmic approximation in $L_1$ norm within 0.08--0.13.  Distances between real distribution excluding language characteristic are in the same interval (Table 3).  90-$\%$ confidence interval is herewith $[0.085; 0.115]$
\par Note that deviations  in norm  $L_1$ of corresponded approximations for both MV-transcriptions are about the same and equal to 0.17. It testifies that a logarithmic model for this text is not adequate enough in case we consider it as an alphabetic cipher from one of any European languages.

\par In order to demonstrate our  language group recognition method, which is based on the statistical analysis of ordered symbols, we consider literary texts in Latin [lat] and Cyrillic [cyr] transcription written  in the following languages: 
\begin{enumerate}
\item Indo-European languages:
\begin{enumerate}
	\item Slavic: East (Russian [rus]), West (Polish [pol], Czech[che]), South (Serbian [serb], Croatian [hr], Bulgarian [bul]);
	\item Germanic: North (Danish[dan], Swedish [swe], Norwegian [nor]), West (German [ger], English [eng], Dutch [hol]);
	\item Romance: Italian [it], Spanish [spa], French [fra], Romanian[rom];
	\item Greek [gre];
	\item Basque [bask];
	\item Latin [lat];
\end{enumerate}
\item Uralic, Finno-Ugric languages:
\begin{enumerate}
	\item Ugric: Hungarian [hung];
	\item Baltic-Finnic: Finnish [fin], Estonian [est];
\end{enumerate}
\item Constructed languages: 
\begin{enumerate}
	\item Esperanto [esp], Volapuk [vol];
	\item Interlingua [int];
	\item Klingon [kl] (language spoken by the Klingons in the Star Trek universe);
	\item Quenya [qu] (``Elvish language").
\end{enumerate}
\end{enumerate}

\begin{table}
\caption{Determination of logarithmic approximation of texts without vocalization for some European languages}
\begin{center}
\begin{tabular}{|g|g|g|g|g|g|}
\hline
[rus-kir]& 0.97 & [ger] & 0.98 & [hr] & 0.91 \\
\hline
[bol-kir]& 0.97 & [eng] & 0.98 & [pol] & 0.96 \\
\hline
[serb-kir]& 0.92 & [hol] & 0.98 & [che] & 0.96 \\
\hline
[gre-kir]& 0.96 & [dan] & 0.93 & [lat] & 0.96 \\
\hline
[fin] & 0.96 & [swe] & 0.96 & [ita] & 0.96\\
\hline
[est] & 0.98 & [nor] & 0.96 & [fra] & 0.96\\
\hline
[hung] & 0.96 & [gre] & 0.96 & [spa] & 0.96\\
\hline
[bask] & 0.96 & [serb] & 0.84 & [rom] & 0.90\\
\hline
\end{tabular}
\end{center}
\end{table}
\par Approximately 90 $\%$ of considered languages have the determination of model logarithmic dependence of frequency odered letters  than  0.96.  Only letter frequency for Danish, Serbian (Latin and Cirillyc alphabets), Croatian and Romanian texts without vocalization have a much lower approximation accuracy.
\par Distances between frequency distributions for texts in Cirillic for Slavic group show that Russian, Bulgarian and Serbian are related: the closest are Russian and Bulgarian (with a distance 0.06), Russian and Serbian as well as Bulgarian and Serbian have a distance 0.12. Note that Greek in Cirillyc transcription have a distance more than 0.20 and in that sense not similar with any of Slavic languages.
\par For texts with the latin alphabet distances between frequency distributions form  clusters (Table 3) in accordance with language groups in sense of closeness between themselves in norm $L_1$. It was found that Indo-European languages  united in groups and subgroups have close statistical properties.  The distance in norm $L_1$ between frequencies from one language group  vary quite narrow (0.08--0.13). Between different groups  the distance is 0.14--0.22
\begin{center}
\begin{table}
\begin{footnotesize}
\caption{{\footnotesize Distance between  frequency distributions  in texts without vocalization (Latin alphabet) in norm $L_1$, $\%$}}
\begin{tabular}{|g|g|g|g|g|g|g|g|g|g|g|g|g|g|g|g|g|g|g|g|g|}
\hline
 & ge & en & ho & da & sw & no & la & it & sp & fr & ro & bs & gr & fi & es & hu & po & ch & hr & se\\
\hline
ge & \cellcolor{darkishgreen}& \cellcolor{RoyalBlue}8 &\cellcolor{RoyalBlue} 11 &\cellcolor{RoyalBlue} 13 &\cellcolor{RoyalBlue} 11 &\cellcolor{RoyalBlue} 12 & 12 & 13 & 11 & 15 & 19 & 18 & 27 & 29 & 14 & \cellcolor{strongRed} 12 & 28 & 26 & 23 & 24\\
\hline
en & &\cellcolor{darkishgreen} &\cellcolor{RoyalBlue} 12 &\cellcolor{RoyalBlue} 13 &\cellcolor{RoyalBlue} 12 &\cellcolor{RoyalBlue} 13 & 12 & 13 & 11 & 15 & 19 & 18 & 26 & 26 & \cellcolor{strongRed} 12 & 16 & 32 & 29 & 28 & 27 \\
\hline
ho & & &\cellcolor{darkishgreen} &\cellcolor{RoyalBlue} 10 &\cellcolor{RoyalBlue} 11 &\cellcolor{RoyalBlue} 11 & 19 & 21 & 19 & 22 & 27 & 25 & 23 & 27 & 20 & 18 & 31 & 28 & 27 & 32\\
\hline
da & & & &\cellcolor{darkishgreen} &\cellcolor{RoyalBlue} 11 &\cellcolor{RoyalBlue} 10 & 13 & 13 & 9 & 14 & 13 & 18 & 27 & 27 & 12 & 16 & 35 & 31 & 30 & 25\\
\hline
sw & & & & &\cellcolor{darkishgreen}&\cellcolor{RoyalBlue} 11 & 15 & 15 & 10 & 14 & 18 & 19 & 26 & 30 & 14 & 13 & 28 & 24 & 23 & 22\\
\hline
no & & & & & &\cellcolor{darkishgreen}& 13 & 13 & 10 & 15 & 14 & 18 & 27 & 27 & 13 & 17 & 34 & 32 & 31 & 26\\ 
\hline
la & & & & & & &\cellcolor{darkishgreen}&\cellcolor{Orange} 5 &\cellcolor{Orange} 10 &\cellcolor{Orange} 7 &\cellcolor{Orange} 12 &\cellcolor{strongRed} 13 & 23 & 22 & 14 & 21 & 39 & 35 & 34 & 32\\
\hline
it & & & & & & & &\cellcolor{darkishgreen}&\cellcolor{Orange} 10 &\cellcolor{Orange} 7 &\cellcolor{Orange} 12 & 15 & 23 & 23 & 14 & 20 & 37 & 35 & 34 & 33\\
\hline
sp & & & & & & & & &\cellcolor{darkishgreen}&\cellcolor{Orange} 11 &\cellcolor{Orange} 13 & 16 & 22 & 25 & 14 & 16 & 36 & 32 & 30 & 28\\
\hline
fr & & & & & & & & & &\cellcolor{darkishgreen}&\cellcolor{Orange} 13 & 16 & 25 & 25 & 18 & 22 & 38 & 35 & 34 & 33\\
\hline
ro & & & & & & & & & & &\cellcolor{darkishgreen}& 20 & 30 & 31 & 21 & 18 & 41 & 38 & 33 & 25\\
\hline
bs & & & & & & & & & & & &\cellcolor{darkishgreen}& 19 & 15 & 15 & 27 & 42 & 39 & 39 & 37\\
\hline
gr & & & & & & & & & & & & &\cellcolor{darkishgreen}&\cellcolor{strongRed} 14 & 21 & 31 & 48 & 45 & 44 & 43\\
\hline
fi & & & & & & & & & & & & & &\cellcolor{darkishgreen}& 22 & 38 & 55 & 52 & 52 & 49\\
\hline
es & & & & & & & & & & & & & & &\cellcolor{darkishgreen}& 18 & 36 & 32 & 31 & 28\\
\hline
hu & & & & & & & & & & & & & & & &\cellcolor{darkishgreen}& 25 & 22 & 17 & 15\\
\hline
po & & & & & & & & & & & & & & & & &\cellcolor{darkishgreen}& \cellcolor{Yellow} 5 &\cellcolor{Yellow} 12 & 22\\
\hline
ch & & & & & & & & & & & & & & & & & &\cellcolor{darkishgreen}& \cellcolor{Yellow} 9 & 20\\
\hline
hr & & & & & & & & & & & & & & & & & & &\cellcolor{darkishgreen}& \cellcolor{strongRed}12\\
\hline
se & & & & & & & & & & & & & & & & & & & &\cellcolor{darkishgreen}\\
\hline
\end{tabular}
\end{footnotesize}
\end{table}
\end{center}
\par The same color in Table 3 is used for language groups were languages are close in pairs in sens of  ordered frequencies of consonant letters in $L_1$. 
\par Red color is used for unexpectedly close language pairs. Clusterization was made on the basis of pair closeness for all units in the cluster. In controversial cases  unit was applied to a cluster with  the largest number of units. 
\par There are some examples of  ordered frequencies of consonant letters in texts in European languages  (Fig. 2-6).

\begin{figure}[ht]
\begin{center}
\begin{minipage}[ht]{0.70\linewidth}
\includegraphics[width=1\linewidth]{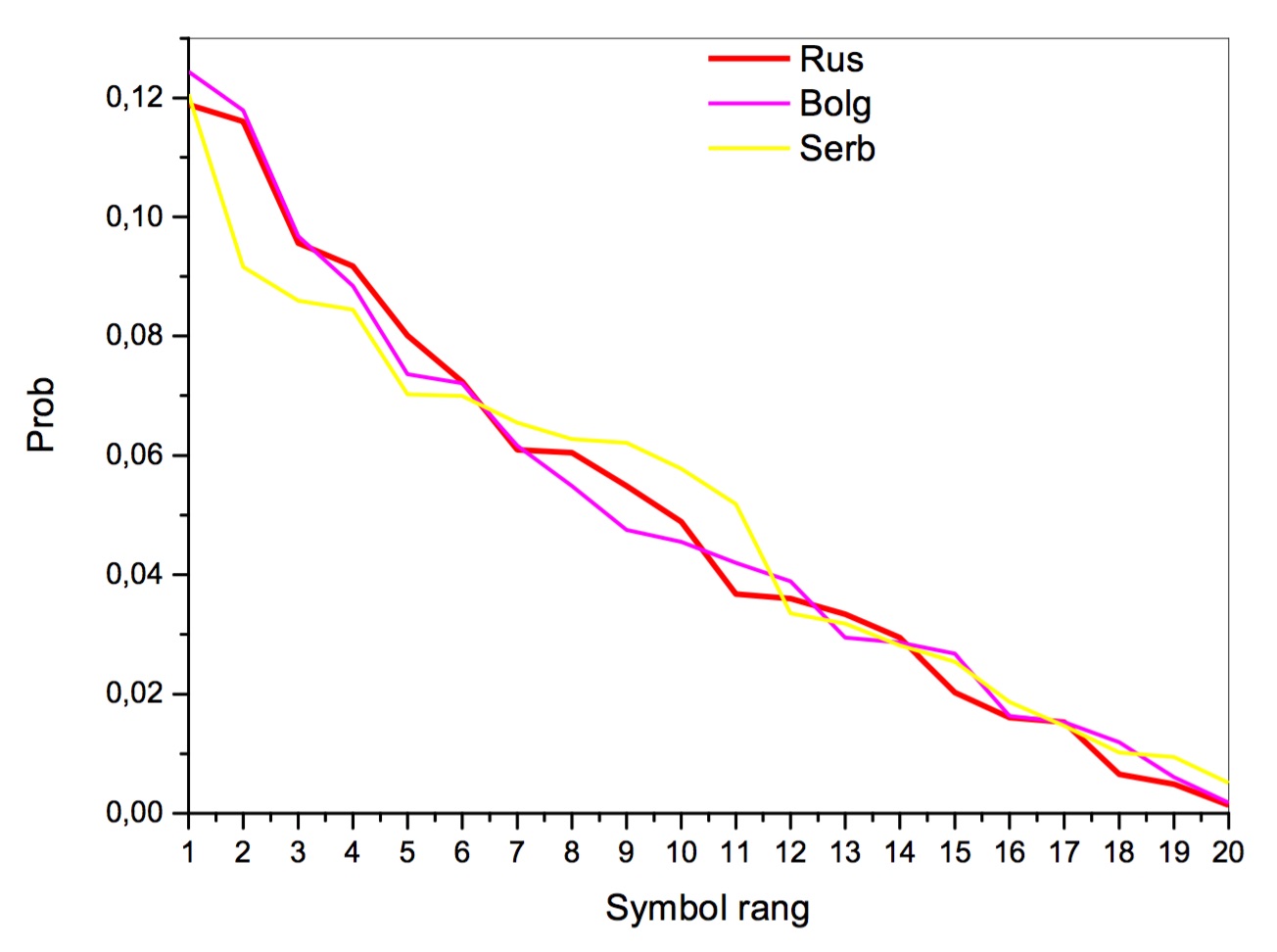}
\vspace{-4mm}
\caption{{\footnotesize Ordered frequencies of consonant letters in texts  (Cyrillic)}}
\label{fig_2}
\end{minipage}
\end{center}
\end{figure}

\begin{figure}[ht]
\begin{center}
\begin{minipage}[ht]{0.70\linewidth}
\includegraphics[width=1\linewidth]{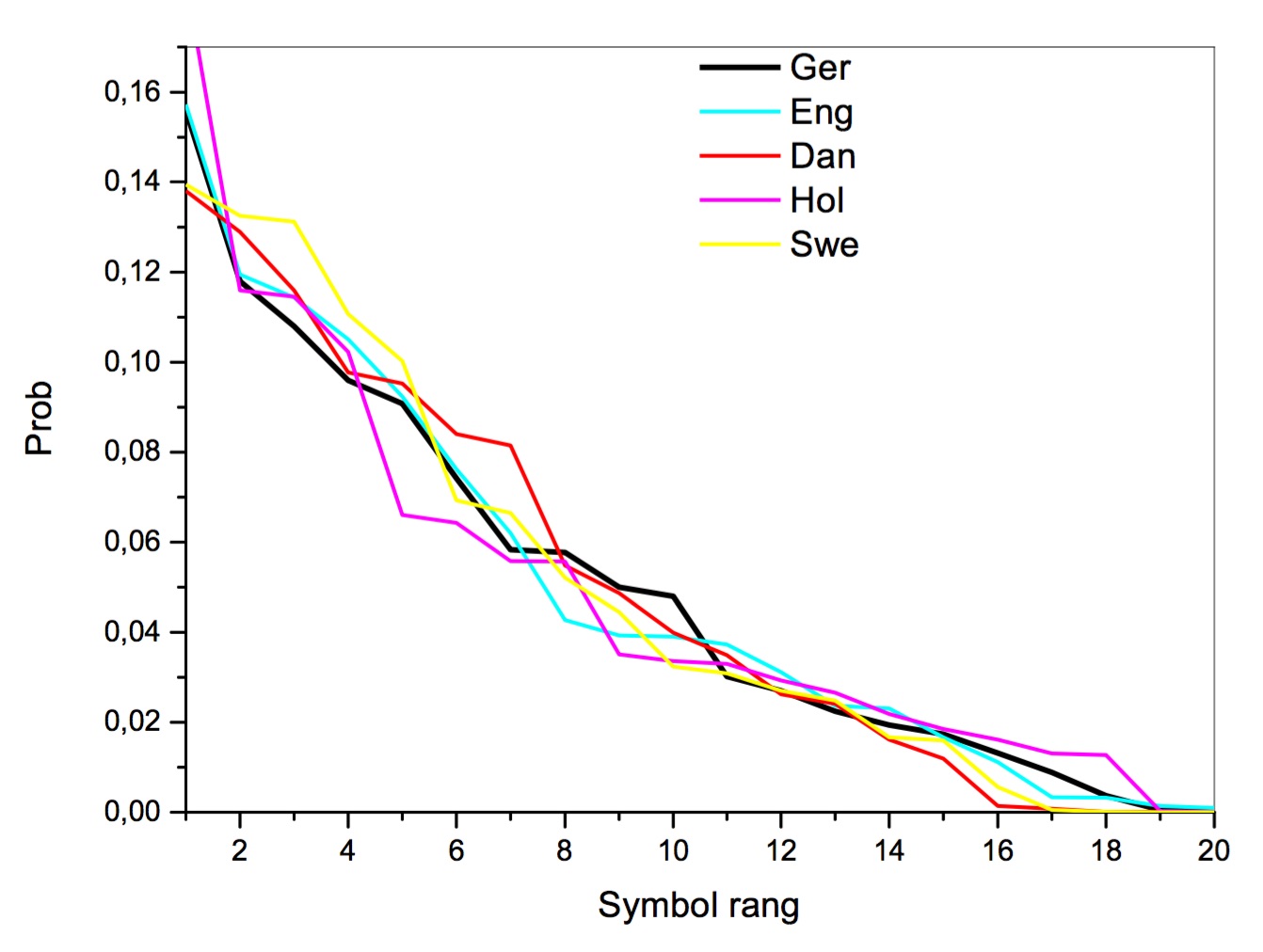}
\vspace{-4mm}
\caption{{\footnotesize Ordered frequencies of consonant letters in texts , Germanic Language}}
\label{fig_3}
\end{minipage}
\end{center}
\end{figure}

\begin{figure}[ht]
\begin{center}
\begin{minipage}[ht]{0.70\linewidth}
\includegraphics[width=1\linewidth]{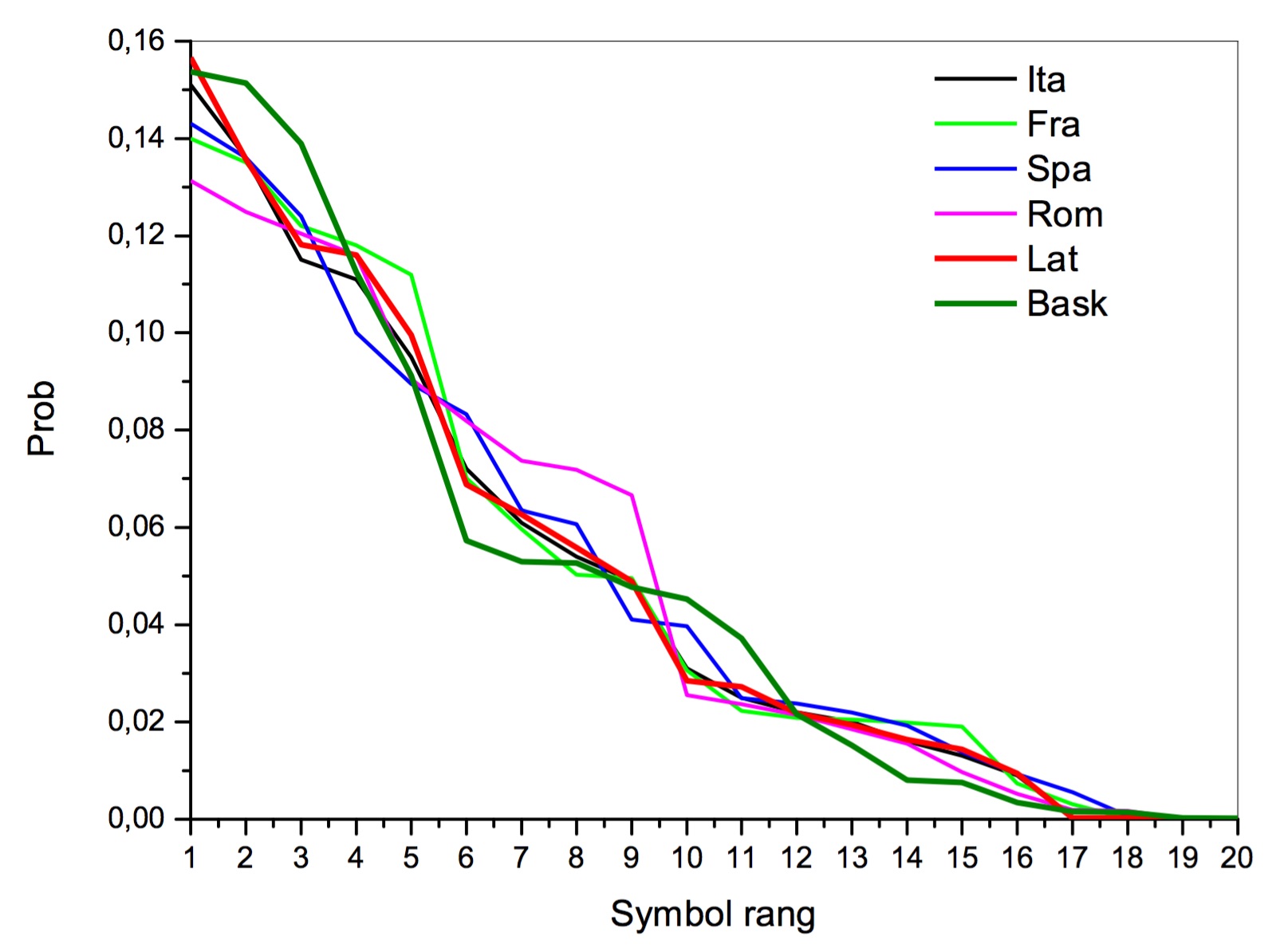}
\vspace{-4mm}
\caption{{\footnotesize Ordered frequencies of consonant letters in texts 
(Romance languages, Latin and Basque)}}
\label{fig_4}
\end{minipage}
\end{center}
\end{figure}

\begin{figure}[ht]
\begin{center}
\begin{minipage}[ht]{0.70\linewidth}
\includegraphics[width=1\linewidth]{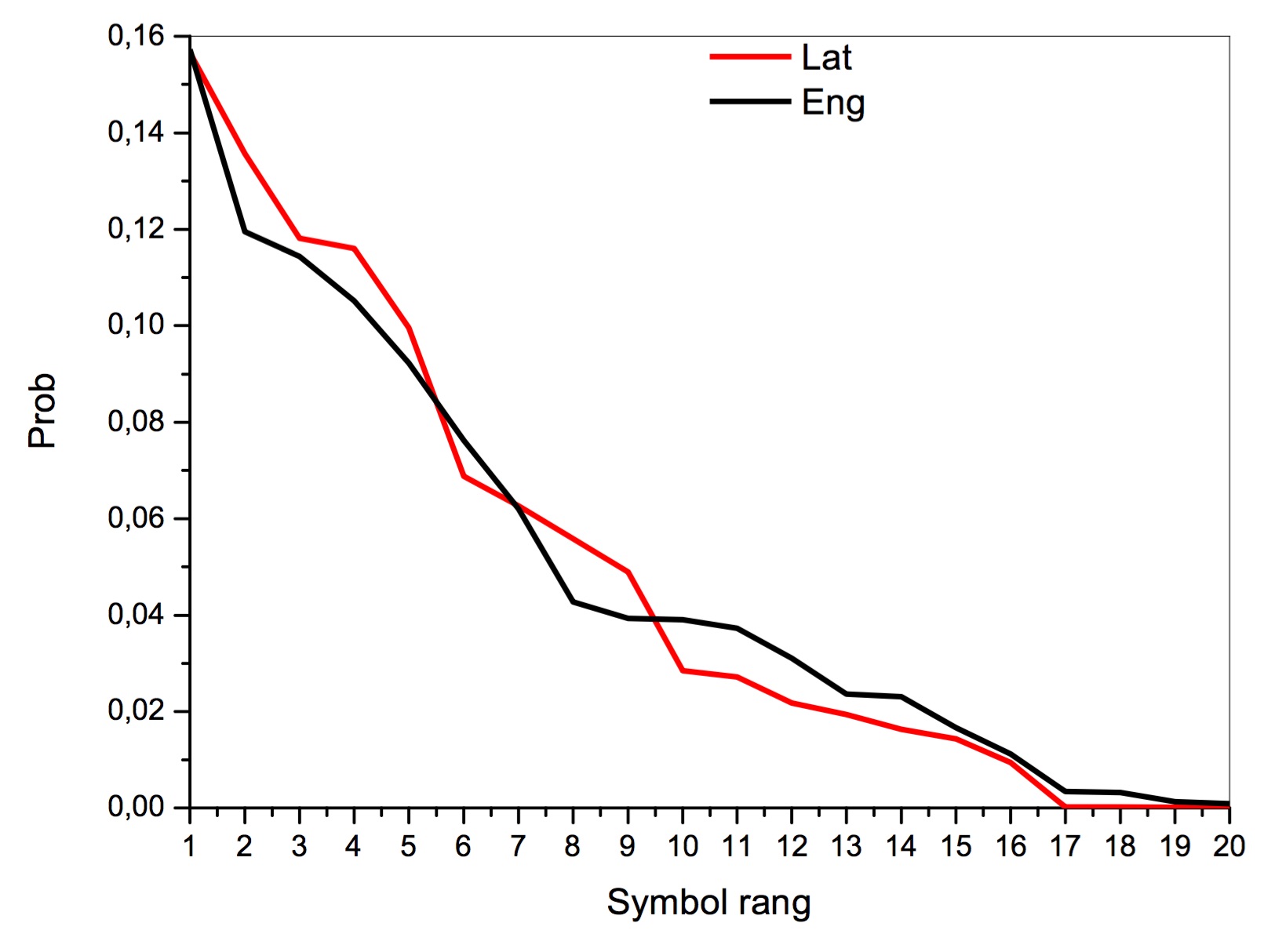}
\vspace{-4mm}
\caption{{\footnotesize Comparison of standard distribution of Germanic and Romance languages}}
\label{fig_5}
\end{minipage}
\end{center}
\end{figure}
\par We have another situation with Uralic family (fig. 6). Frequencies of consonant letters for Finnish and Estonian languages are markedly different (they have the distance 0.22) while they are both in Baltic-Finnic branch. The line corresponded with Hungarian has a distance from these languages 0.38 and 0.18 respectively and is closer to Germanic languages with distance 0.16. This result is probably caused by the fact that vowels have more significant structure value in these languages. 

\begin{figure}[ht]
\begin{center}
\begin{minipage}[ht]{0.70\linewidth}
\includegraphics[width=1\linewidth]{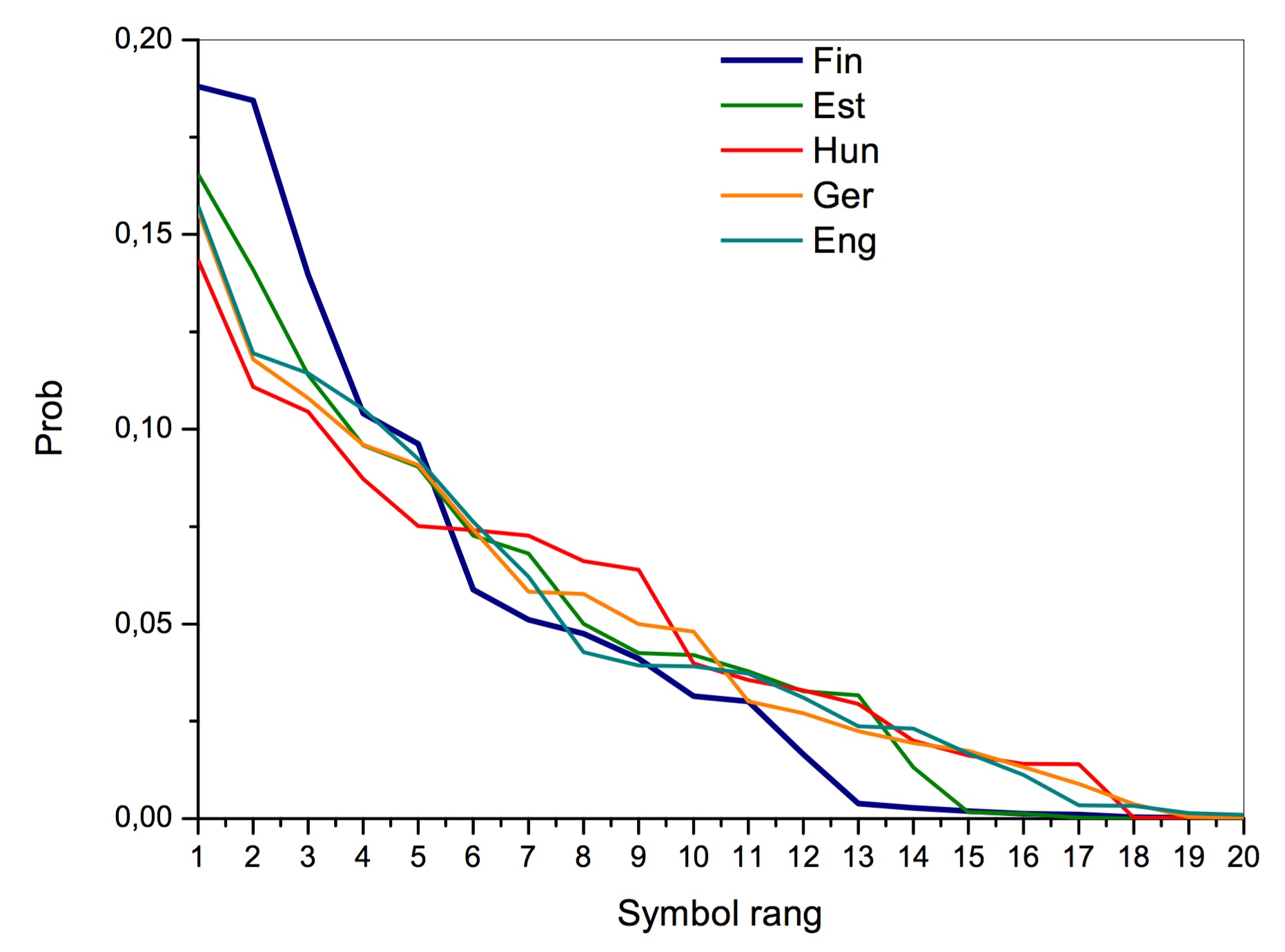}
\vspace{-4mm}
\caption{{\footnotesize Ordered frequencies of consonant letters in texts (Uralic vs Germanic)}}
\label{fig_6}
\end{minipage}
\end{center}
\end{figure}
\par Note that the Latin alphabet is used in Southeast Asia, for example in Vietnam.  However, frequency of consonant letters in Vietnamese texts  has a distance from  European texts more than 0.25. Vietnamese is not united with any of European languages from Indo-European or Uralic families. MV language is also not similar with Vietnamese. Virtually, Latin used in Vietnamese texts (without diacritical marks)  has a minimum distance 0.28 (in Takahashi transcription) from VM. Latin without vocalization  has a distance even more --  0.35. In this way ``Vietnamese version" about VM is not confirmed.
\par Constructed language texts will be discussed in a later section.
The results can be commented by these allegations:
\begin{itemize}
\item Languages, which are linguistically attributable to one group, for the most part are in the same cluster with a distance less than 0.13. This applies to German, Romance and Slavic groups.
\item Greek language in the Latin transcription is aside from the rest of the languages, as, indeed, Finnish language, but to each other they were closer than each of them to other languages. The distance between them was only 0.14, which marked in red in the table. 3.
\item Estonian and Finnish languages obviously have different statistical properties    though they are united in the same group by linguists. Statistically, Estonian and Hungarian are closer to the German group, the distance to which they amounted to 0.12--0.14.
\item Latin transcription of the Serbian language is only close to the Croatian language with 0.12 distance between them.
\item Slavic group of languages (Czech, Polish and Croatian), that use the Latin alphabet, are tightly clustered.
\item Although the Medieval Latin is out of use, it is close to all languages in the Romance group. Basque language, who has an uncertain status, is away from all the languages in question except from the Latin, to which the distance was only 0.13. Both languages have the same determination of the logarithmic model (0.96) with the same number of effectively used consonants (16).
\end{itemize}
\par Thus, clustering method on the principle of the pairwise proximity between distributions allowed to obtain meaningful results of linguistic classification of languages. As mentioned above, these distributions have high logarithmic profile accuracy.
\par The logarithmic model of symbol distribution has been derived by S. M. Gusein-Zade \cite{Gusein}. It is based on the assumption of a constant distribution density of the random point $P(p_1,....,p_n)$ on $n$-dimensional simplex $\sum\limits_{i=1}^{n} p_i=1$, where $p_i$ is the frequency of use the order of $i$ letter in the text. In \cite{OrOs} this model was modified and used to evaluate the completeness of the alphabet in the texts in different languages. It represents 
\begin{equation}\label{1}
f(k)=\frac{1}{n}  \left( 1+\frac{1}{n+o}\ln{\frac{n!}{k^n}} \right)
\end{equation} 
\par In this formula $n$ is the number of letters in the alphabet. The parameter $o$ is the nearest integer, corresponding to the value with the smallest error of approximation of the actual distribution of the formula (\ref{1}). The point of this parameter consists that the text under consideration has the most adequate text alphabet with $n+o$ the number of symbols. Empirical relationships for Russian, German, English and Hungarian languages in Fig. 2 are best modeled by relation (\ref{1}) wherein $o=0$; this option is shown in Fig. 1 as legend {\it model}. For French, Spanish and Italian, Danish and Swedish $o=-1$. For Finnish $o=-6$ and for Estonian language $o=-4$.
\par Applied to EVA transcription, for which $n = 22$, the best approximation is achieved when $o = -2$. This means that in MV has only 20 symbol in use. The elimination two of the rarest letters obtains a logarithmic approximation with 0.93 determination and a deviation in the $L_1$ norm of the actual distribution at the level of 0.167. The same observation is true for Takahashi transcription. Appropriate depending were presented in Fig. 1.
\par Observe now that the number of consonants in most European languages is 20. It could be assumed that the MV is written on one of them, but without vowels. Necessary in a statistical sense, but not a sufficient condition for this is, firstly, the proximity of one of the transcriptions' distributions to the selected language distribution (deviation in the $L_1$ norm does not exceed 0.10). Secondly, it has to be a roughly equal distance from transcription and from selected language to approximating model dependence (about 0.17). Analysis of the data in Fig. 3, 4, 6, lead to the conclusion that from the considered options there is only one suitable -- Danish language. Takahashi Transcription has 0.10 divergence in $L_1$ norm from the empirical distribution of the Danish language (Fig. 7).

\par What is more, the determination of the logarithmic approximation of Danish language without vowels is 0.93, the same to Takahashi transcription. Swedish and Norwegian (Bokmal) languages that close to the Danish are much less suited to the role of the language of the original MV. Distances between all the languages of the North German group are the same and is equal to 0.11 (differences are only in the third decimal place), and the difference between the Swedish and Norwegian from this transcription is 0.14 instead of 0.10 for the Danish language.

\par It is worth noting that Latin can be considered as the language of MV because a deviation Takahashi transcription in this case is 0.11, which is also within the allowable distance between distributions of related language groups.

\begin{figure}[ht]
\begin{center}
\begin{minipage}[ht]{0.70\linewidth}
\includegraphics[width=1\linewidth]{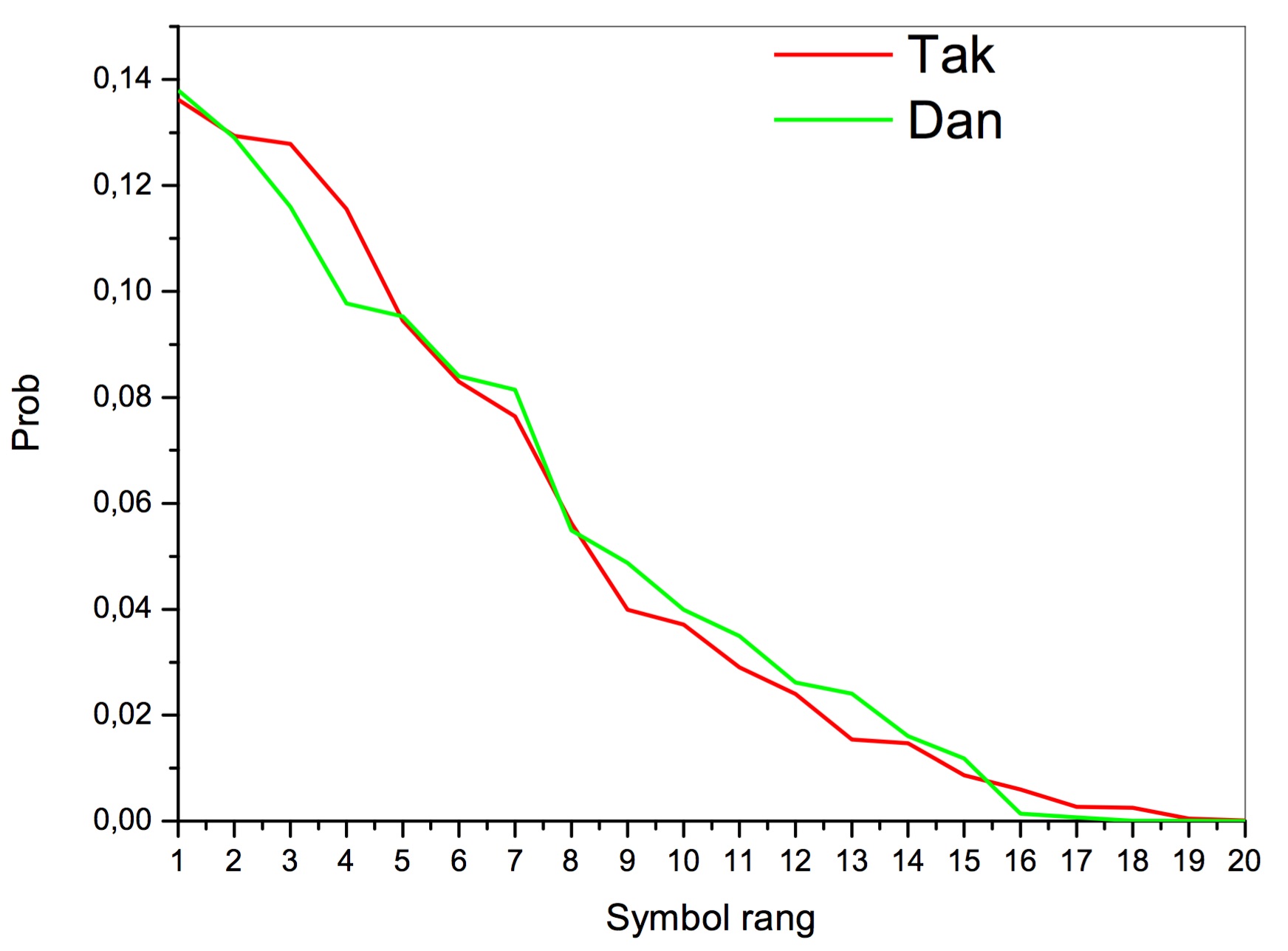}
\vspace{-4mm}
\caption{{\footnotesize Distribution of symbols frequencies in Takahashi transcription and Danish texts without vowels}}
\label{fig_7}
\end{minipage}
\end{center}
\end{figure}

\par For EVA transcription a suitable language among discussed European was not found.
\par The findings are based on a statistical analysis of modern texts. Transferring selected properties on manuscripts XVI-XVII centuries can be made only under the assumption that the use of the lexicon consonants during that time has not changed significantly. Unfortunately, the number of the original texts were not available in sufficient amount to analysis, so the expressed hypothesis is only illustrates the method of analysis of texts, and is very preliminary. However, it should be noted that the analysis of the texts in Medieval Latin led to a similar dependence: 0.966 determination of the logarithmic model without vowels. This finding suggests that the logarithmic model in some sense is invariant, and can be used to compare the distributions of symbol in both old and modern texts.
\par It has to be emphasized that conjectural identification of the language does not mean readability of the manuscript, since an invariant of language is the curve, rather than a position specific letters on it. In different texts in the same language the ordered sequence of letters is  ``floating", although, the most frequently used symbols in one text are not become rarely used in another. Nevertheless in any language there is no unambiguous correspondence between the rank and the letter.  The width of ``migration window" rank for text on the length of the order of 200 thousand signs is 5. Therefore, even if the original language manuscripts has been specified, it cannot guarantee the transcript. Extra analysis on a large set of texts is required in order to establish the possible combinations of letters in their orderly distribution. 

\par One argument in favor to fact that the manuscript is written in a language without vowels was given. However, considering the Danish language, the possibility that MV was written in language with a full alphabet, but excluding diacritical marks, by which North German languages (and not only) group are saturated with , should take into account. At the same time the number of actually used consonants may be less than 20, as the letters Q, X, W, Z are used in Danish only in borrowed words, which could simply not to be in a manuscript. Then, the number of different symbols would be 22, but for other reasons. However, reviewed below analysis of other statistics still evidence in favor of concept of the manuscript as ``with vowels excluded".

\section {Distribution of distances among pairs of the identical letters}
\par In this section we will treat a text as a time series with values $x(t)$, where $t$ is the ordinal number of symbol $x$ counted from the beginning, and $x$ itself belongs to some fixed alphabet. The length of a symbol sequence, which do not contain particular letter, is an important property of the underlying language, as its distribution does not change remarkably between various texts.
In particular, it was shown in \cite{OrOs} that for Russian texts distances between the same symbols, i.e. the number of all other letters between them, have the following properties:
\begin{itemize}
\item[$-$] Autocorrelation function is near zero for every value of time lag, but increments are dependent;
distribution of distances is preserved across texts of different authors and styles;
\item[$-$] After multiplication by normalization constant equal to the frequency of the symbol occurrence this distribution does not depend on the symbol.
\end{itemize}

\par It should be noted that symbols in a text does not appear spontaneously, but in accordance with the narration and author's conceptions. Hence distances between the same letters cannot be purely random, and there could be a parameter governing the effects of long-range memory in symbol appearance. As it was mentioned in \cite{ZKO}, the order of derivatives in the fractional Fokker--Planck equation for empirical probability densities of arbitrary time series might play the same role.
\par Let us analyze numerical time series of distances between some predefined symbols. For the sake of computational simplicity we will use the well-known Hurst exponent \cite{KKMP} instead of the fractional derivative statistics from \cite{ZKO}. A value of the Hurst exponent agrees well with fractional derivative order in case of self-similar time series.
\par The Hurst exponent can be estimated as follows. 
\par Procedure is applied to the time series formed by first differences $x(t)= b(t+1)-b(t)$ of the original time series $b(t)$. Firstly, one calculates the moving average values and constructs new series of accumulated deviations from the moving average for sub-frames of length $k$:
$$
\overline{x}(t,k)=\frac{1}{k}\sum\limits_{i=t-k+1}^{t}x(i)
$$
\par Then the range, which is defined as the difference between the maximum and the minimum values of this auxiliary series, and the standard deviation of the original time series $b(t)$. Firstly, one calculates the moving average values and constructs new series of accumulated deviations from the moving average for sub-frames of length $k$:
\begin{equation}\label{2}
R(t,k)= \max_{j\le t} \left(\sum\limits_{t-k+1}^{t}(x(i)-\overline{x}(t,k)) \right)-\min_{j\le t}\left(\sum\limits_{t-k+1}^{j}(x(i)-\overline{x}(t,k)) \right)
\end{equation}

$$
\sigma_{x}^{2} (t,k)= \frac{1}{k}\sum\limits_{t-k+1}^{j}{(x(i)-\overline{x}(t,k))}^{2}
$$
Each range is divided by the corresponding standard deviation, then the arithmetic mean of logarithms of rescaled ranges is calculated:
$$
\xi (t,k)=\ln ({R(t,k)\over \sigma_{x}(t,k)}), \quad \overline{\xi}_N(t)={1\over {N}}\sum\limits_{k=1}^{N} \xi (t,k)
$$
The Hurst exponent $H_{N}(t)$ at time-step $t$ is estimated as the slope in linear regression model of $\xi(t, k)$ on logarithms of $k$:
\begin{equation}\label{3}
H_{N}(t)={1\over {N}} \sum\limits_{k=1}^{N} \left( \xi (t,k)- \overline{\xi}_{N}(t)\right)\left(1+\ln(k/N)\right)
\end{equation}
\par As calculations showed, distances between identical symbols for all considered languages, independently of the presence or absence of vowel letters, form the so-called antipersistent time series, because their Hurst exponents are significantly lower than 0.5, which corresponds to the standard Wiener process. Empirical distributions of Hurst exponents obtained by the rescaled range technique for several languages with $N = 5000$ are depicted on Fig. 8.
\par In the figure ``eng-tot'' denotes the distribution for an English text written in the full alphabet, and ``eng-consonant'' corresponds to the same text without vowel letters. As one can see, distributions for Russian and English have sharp maximums in different locations. Probably, distribution of Hurst exponents can be used to identify a language (or at least language group) of a text, but this topic is beyond the scope of the present paper.
\par Now it can be stated only that our preliminary conjecture made in section 1 about Danish (or Spanish) as the original language of the Manuscript should be discarded. The reason is that the distributions of Hurst exponents for the Manuscript and ordinary texts are completely different. In case of the Manuscript observed distributions are shifted to the right and have much less acute maximum compared to all other curves on Fig.8. This means that statistics of the Manuscript does not agree with statistics of texts written in one particular language. Roughly speaking, symbols in the Manuscript are placed ``more randomly" compared to the latter. Further analysis of these issues will be presented in the following sections of the paper. There are two main options here: the Manuscript is written in a special constructed language or it is written in several languages.

\begin{figure}[h]
\begin{center}
\begin{minipage}[ht]{0.70\linewidth}
\includegraphics[width=1\linewidth]{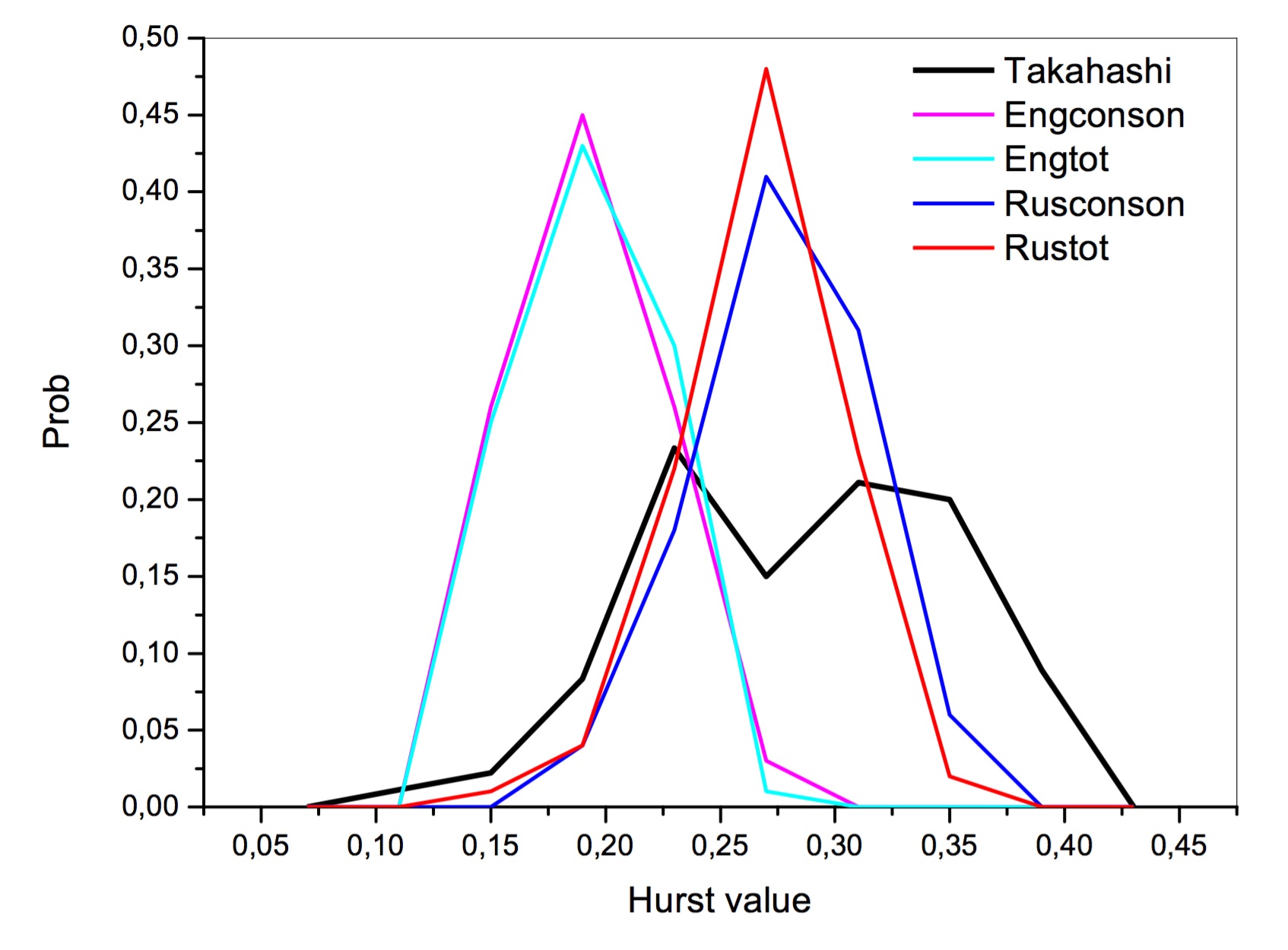}
\vspace{-4mm}
\caption{{\footnotesize Empirical distributions of Hurst exponents for time series formed by distances between pairs of the most frequent letters for several languages}}
\label{fig_8}
\end{minipage}
\end{center}
\end{figure}

It should be pointed out that in the case of multilingual texts (i.e. several languages are mixed together) the Hurst exponent is ineffective as a language indicator, because one does not know in advance which parts are written in each language. Distributions of Hurst exponents are almost similar for texts with and without vowels, hence they cannot indicate presence or absence of the latter. Nevertheless, provided that the distribution of Hurst exponents is clearly unimodal, one can state that the text is written in one particular language.

\section {Constructed languages symbols statistics}
\par One of explanation for the deviation statistics of symbols of Manuscript from the statistics of ``ordinary" language can consist in the fact, that the Manuscript language could be constructed. There are about hundreds of constructed languages. Most of them are built on a variety of ideas, combining natural languages however there are fiction languages in literary works. These languages, however, are not very different from the natural in terms of their statistical properties. Fig. 9--10 shows the distributions for some of them.

\begin{figure}[t]
\begin{multicols}{2}
\hfill
\includegraphics[width=70mm]{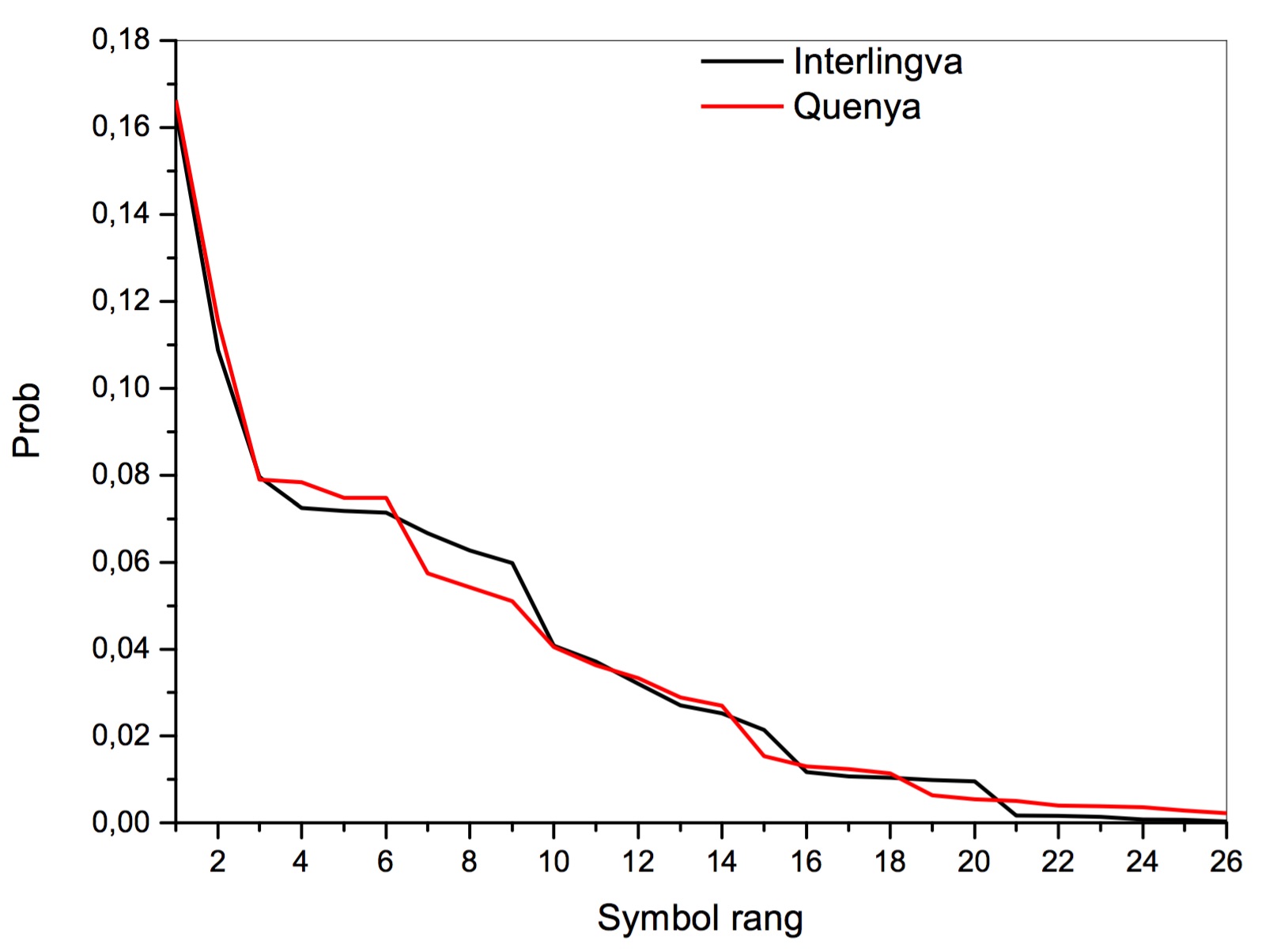}
\hfill
\caption{{\footnotesize Distribution of symbols of ordering some constructed languages}}
\label{fig__9}
\hfill
\includegraphics[width=70mm]{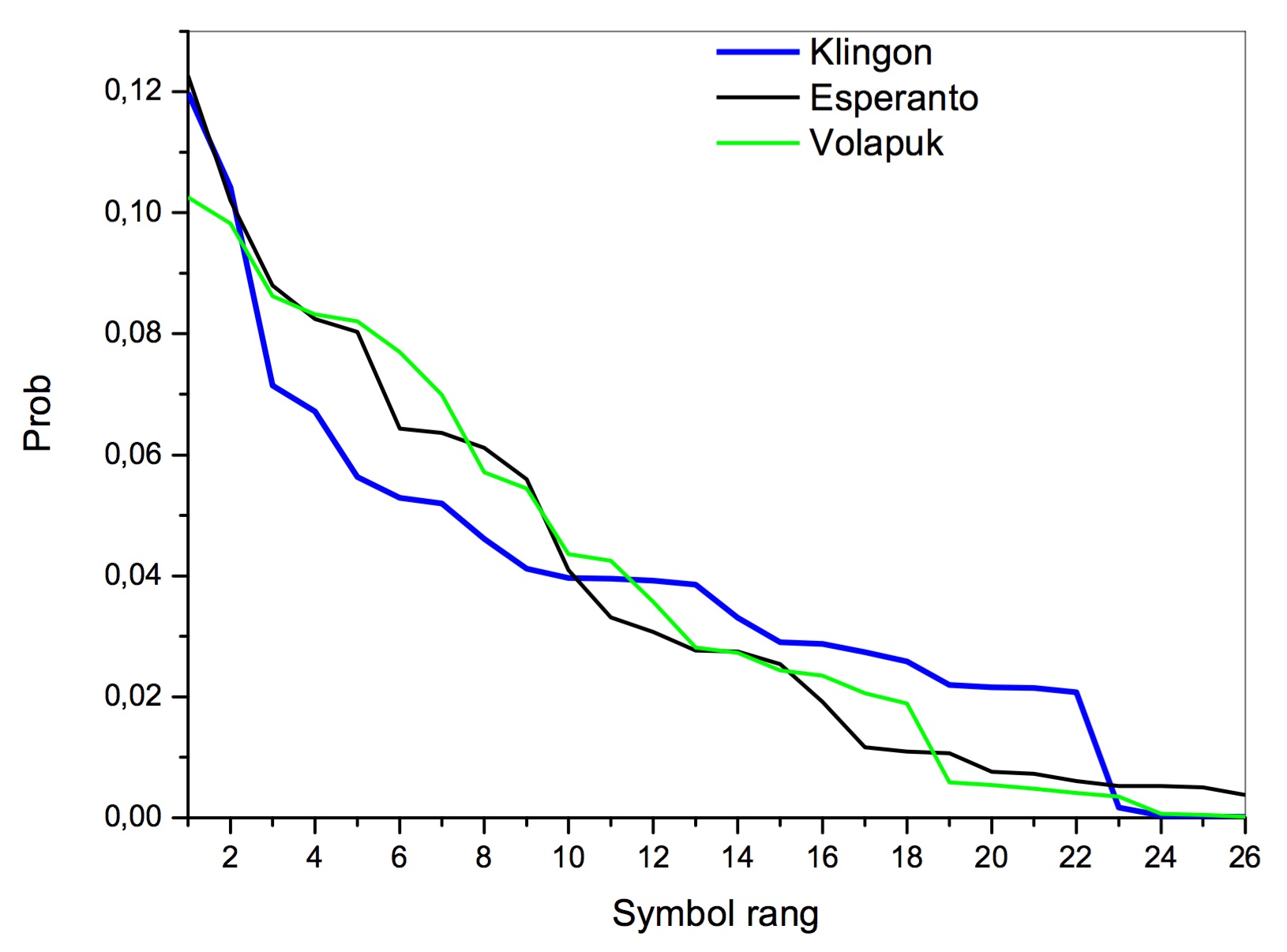}
\hfill
\caption{{\footnotesize Distribution of symbols of ordering some constructed languages}}
\label{fig__10}
\end{multicols}
\end{figure}

\par It appeared that the ``Elvish" language Quenya is very close to interlingua language and Esperanto and Volapuk close to Latin, Klingon language stands alone. Distances between all these distributions is relatively small -- from 0.09 to 0.13, but it must be noted that in this example, the complete alphabets were considered. Their distances vary from 0.03 to 0.05 in the case of close natural languages.
\par Observe now that the same factor of the presence of chains of three identical symbol in a row can be interpreted both in terms of reading the without vowel MV and as part of hypothesis of an constructed language manuscript, where symbols can have a syntactic role. In this meaning the text in Esperanto is very close to EVA transcription (Fig. 11). The distance between them in $L_1$ norm was only 0.11, with the main difference in the low frequencies, rather than larger ones.
\par The reliability of the logarithmic approximation in constructed languages is somewhat worse than in the texts on Indo-European languages, and stays at the level of determination of ``without vowels" texts. The determination of the Klingon language is 0.95; Elvish -- 0.96; Esperanto -- 0.97. This does not mean that any constructed language has the same high determination, but all the same determination MV significantly below these values.

\par This work is not aimed to highlight detailed analysis of constructed languages. The purpose was to give some particular demonstration of an assumptions that statistical properties similar to properties of natural languages are expected from the language which is sufficient to write meaningful text of a large size.
\par Fig. 12 shows the distribution of the Hurst exponent for rows of distances between the same symbols for constructed languages. The graph of the distribution of the Hurst exponent for MV in Takahashi transcription is also presented.
\begin{figure}
\begin{center}
\begin{minipage}[h]{0.70\linewidth}
\includegraphics[width=1\linewidth]{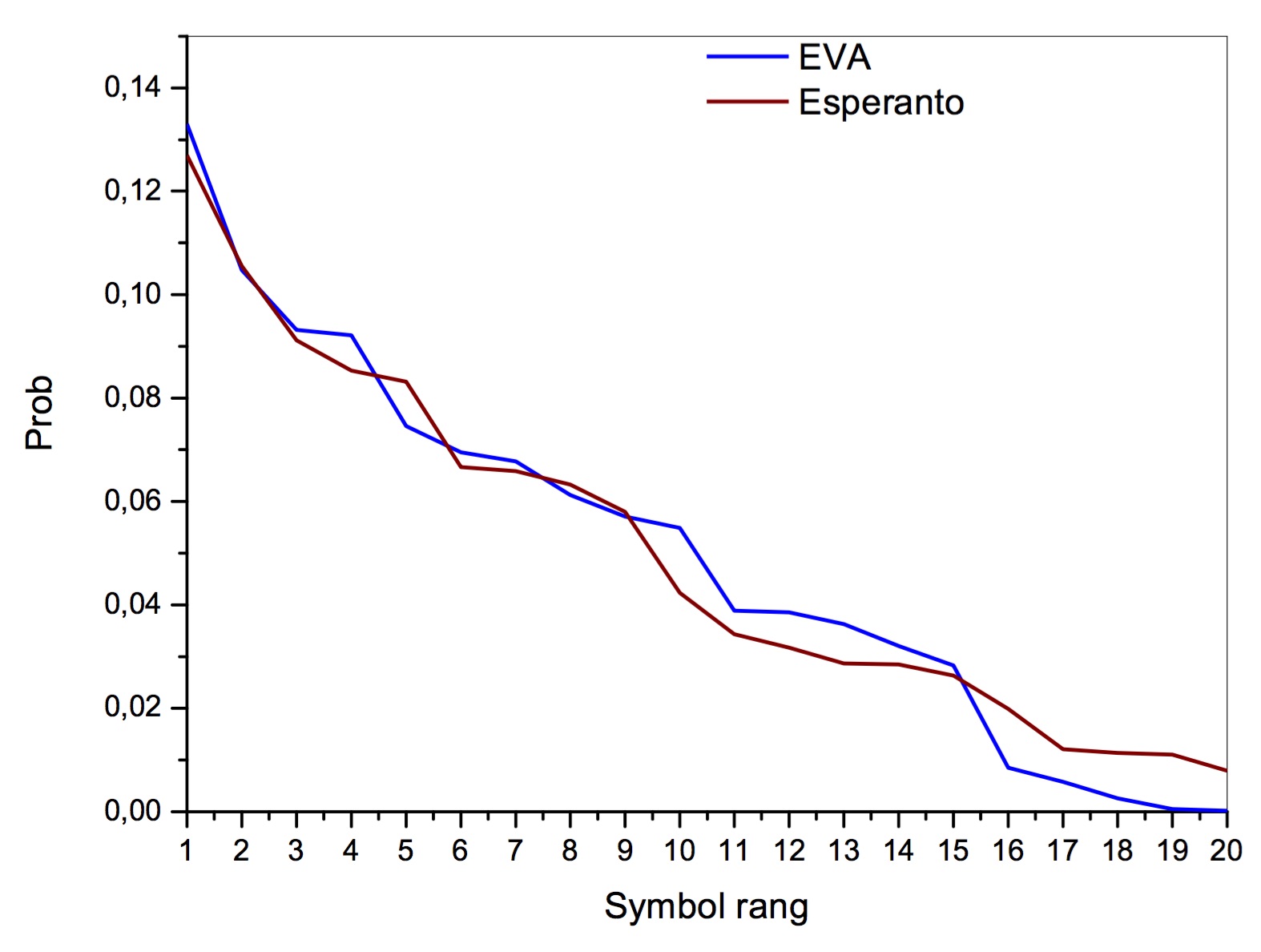}
\vspace{-4mm}
\caption{{\footnotesize Distribution of symbol by ordering text in Esperanto and EVA transcription}}
\label{fig_11}
\end{minipage}
\end{center}
\end{figure}

\begin{figure}
\begin{center}
\begin{minipage}[H]{0.70\linewidth}
\includegraphics[width=1\linewidth]{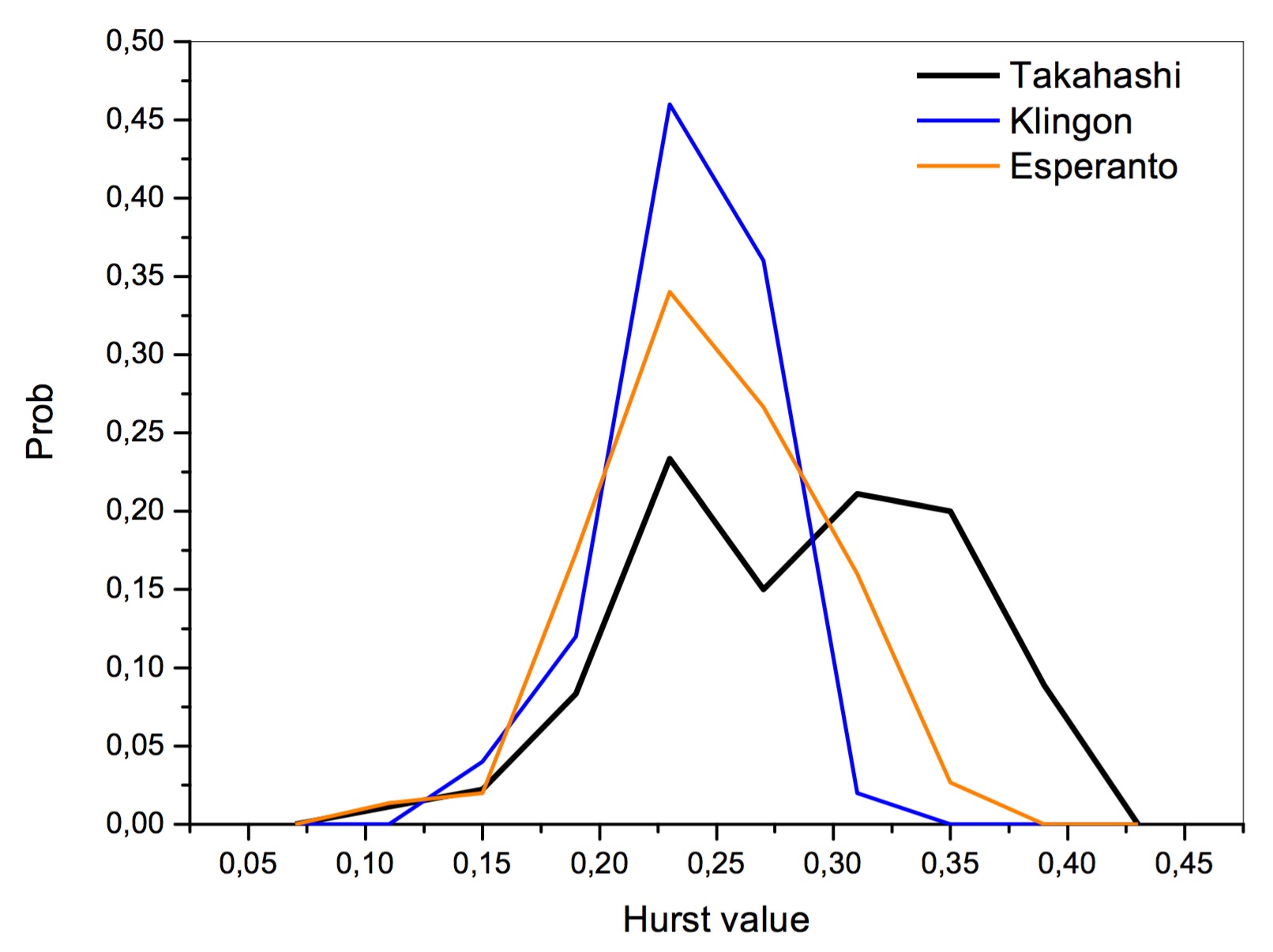}
\vspace{-4mm}
\caption{{\footnotesize Distribution of the Hurst exponent for the rows of distances of between same letters in the texts on the constructed languages}}
\label{fig_12}
\end{minipage}
\end{center}
\end{figure}
\par The fact that Hurst exponent for texts on natural and constructed languages behave roughly the same, and this behavior is different from the MV can be observed from Fig.8 and Fig.12. It seems that for the case of a sufficient elaboration of constructed languages they have antipersistent behavior of a number of distances between same letters in the texts. Consequently, the option of constructed language of the manuscript at this stage should be excluded.
\par As already mentioned above, more acclivous shape of the Hurst exponent for MV and the absence of its unimodal distribution can be the result of mixing several languages in the text, in particular two languages.
\section {Two-languages symbols statistics}

\medskip The version of the language, that we found in Sec.1 and in which the Manuscript could have been written, doesn't cover other ways to build a text with a symbol statistics close to Takahashi transcription. The deviation of the VM logarithmic approximation of frequencies from an actual distribution approximately equals to 0.17, which is substantially more than the deviation from the majority of texts in European languages. This fact indicates that VM might have been written in two languages with a common alphabet. The last condition is not required but simplifies our research. 
\par Non-vowels text determination, observed in most European languages at 0.96 rate, can be decreased to 0.93, which is similar to VM determination, assuming a text is bilingual, written in two languages with a common alphabet, e.g. roman. After the  removal of vowels and decoding this text turns into so-called Voynich Manuscript. We expect that similar letters in both languages are not designated as different symbols in the Manuscript, which certainly narrows the search area. Still, it is worth mentioning that the probability of different alphabets usage is low, as Takahashi transcription determination is greater than 0.9. For such usage it is necessary to know symbol frequencies in each alphabet and group reassigned symbols properly, which seems to be rather unrealistic for the XVI century, especially considering the fact that regression analysis was invented much later. For this reason one should assume that the Manuscript text is meaningful, otherwise a deviation from specific letter statistics for the natural lexicon would have been much greater.
Thus, in this section we accept the following working hypotheses regarding the VM:
\begin{enumerate}
\item The manuscript is a bilingual text with a common alphabet.
\item Vowels have been deleted from the text before the decoding.
\item Decoding was a bijective letter replacing by a symbol
\item Spaces in the text are not considered as characters
\end{enumerate}
\par Then we need to find out which pairs of languages with a common alphabet and in which proportion could be considered as the Manuscript languages, whether they have the same or different linguistic groups and which goups exactly. In addition to that, we need to discover how much a thematic aspect affects the statistical properties of the texts. A genre influence on alphabetic (not frequency-ordered) distribution in Russian texts  was considered in \cite{OO}, where a definite relationship has been observed.
We will show here the results of statistical analysis of the frequency in modern texts written in two different languages but with the same alphabets. It should be done in any case to test the hypothesis of the model determination reduction (\ref{1}) by mixing texts languages. To test this hypothesis, we join two texts with about equal volumes, each one written in its own language but with the same alphabets in both texts. We will analyze texts without vowels.
\par We consider texts with the same language group first.
\par Russian and Bulgarian non-vowel texts distribution average can be seen in Fig.13. Also there is a model dependency graph (\ref{1}) for 20-letters alphabet for zero value of the parameter . It was found that pure and 50/50 Russian and Bulgarian mixed texts both have similar distributions with the similar determination, which equals to 0.96, and an actual distribution deviation from the model, which is equal to 0.10. This mixture, obviously, has different statistical properties from the VM in either of the two transcriptions.
\begin{figure}
\begin{center}
\begin{minipage}[H]{0.70\linewidth}
\includegraphics[width=1\linewidth]{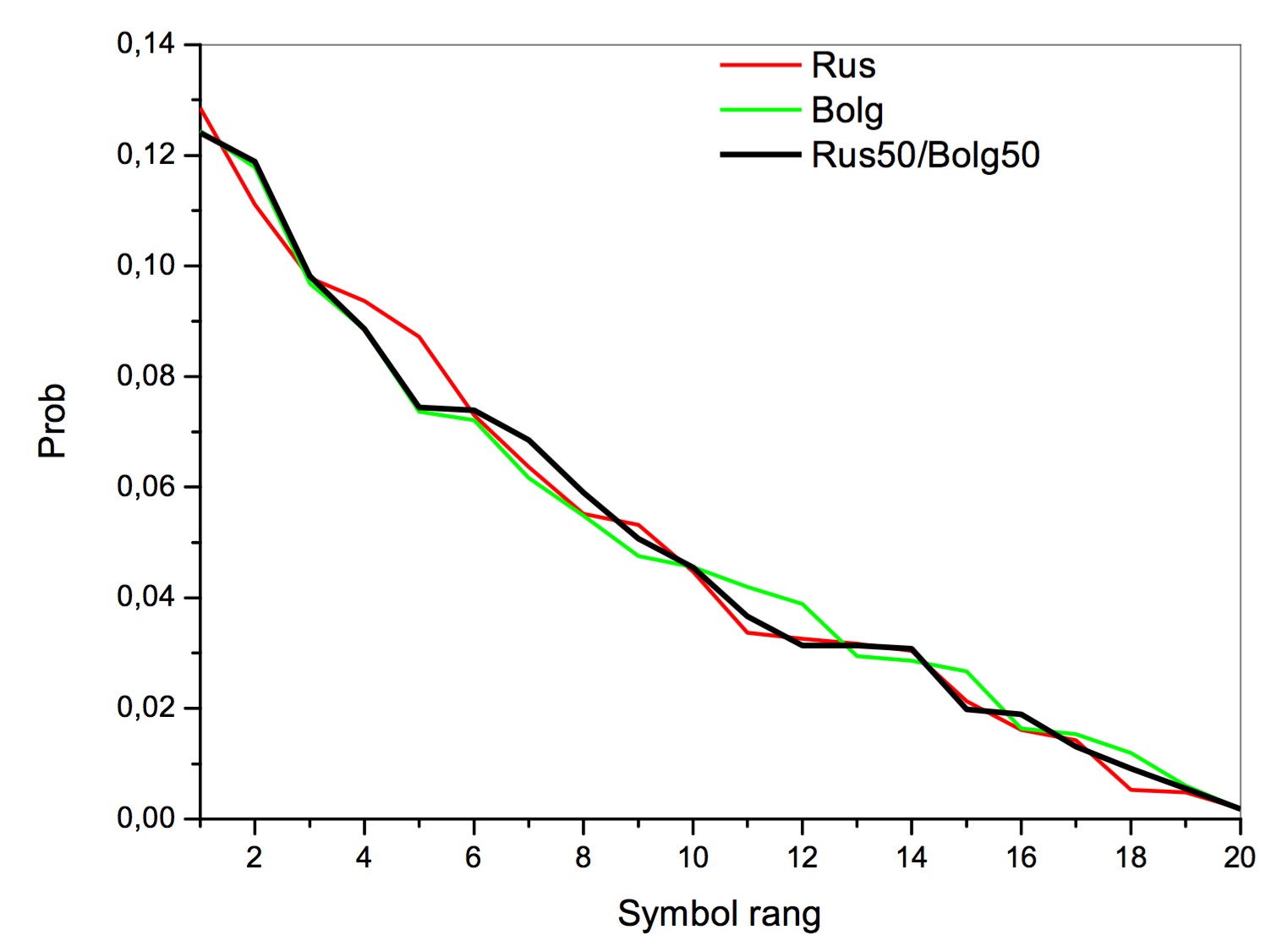}
\vspace{-4mm}
\caption{{\footnotesize The mixture of Russian $\&$ Bulgarian Cyrillic texts}}
\label{fig_13}
\end{minipage}
\end{center}
\end{figure}
\par Similar distributions for English-German texts are shown in the Fig.14. For these language pair as well as for French/Italian pair and for all text with the same language groups or subgroups (among the following three groups: Slavic, Germanic and Romance) mixture determination of logarithmic approximation coincides with the determination of texts in one language at the same rate of 0.96.
\par Thus, languages with the same group not only have close ordered frequency distribution in the texts without vowels, but also a mixture of these languages has the same logarithmic approximation determination to its components. It might to be interesting to check this observation against the texts of the XVI century, but in this paper we stick to the analysis of modern texts statistics only, realizing that the findings are not strictly conclusive with respect to the VM.
\par We now consider examples of texts mixture of a different language group of Indo-European family. In the Fig.15 the frequency distributions in Spanish-English texts without vowels are shown. Notice that a mixture of equal proportions of English and Spanish leads to a determination at 0.92 with a deviation of 0.17 from an actual distribution in the $L_1$-norm approximation. According to the statistics this mixture looks like Takahashi transcription, the distance between two distributions is 0.12.

\begin{figure}
\begin{multicols}{2}
\hfill
\includegraphics[width=70mm]{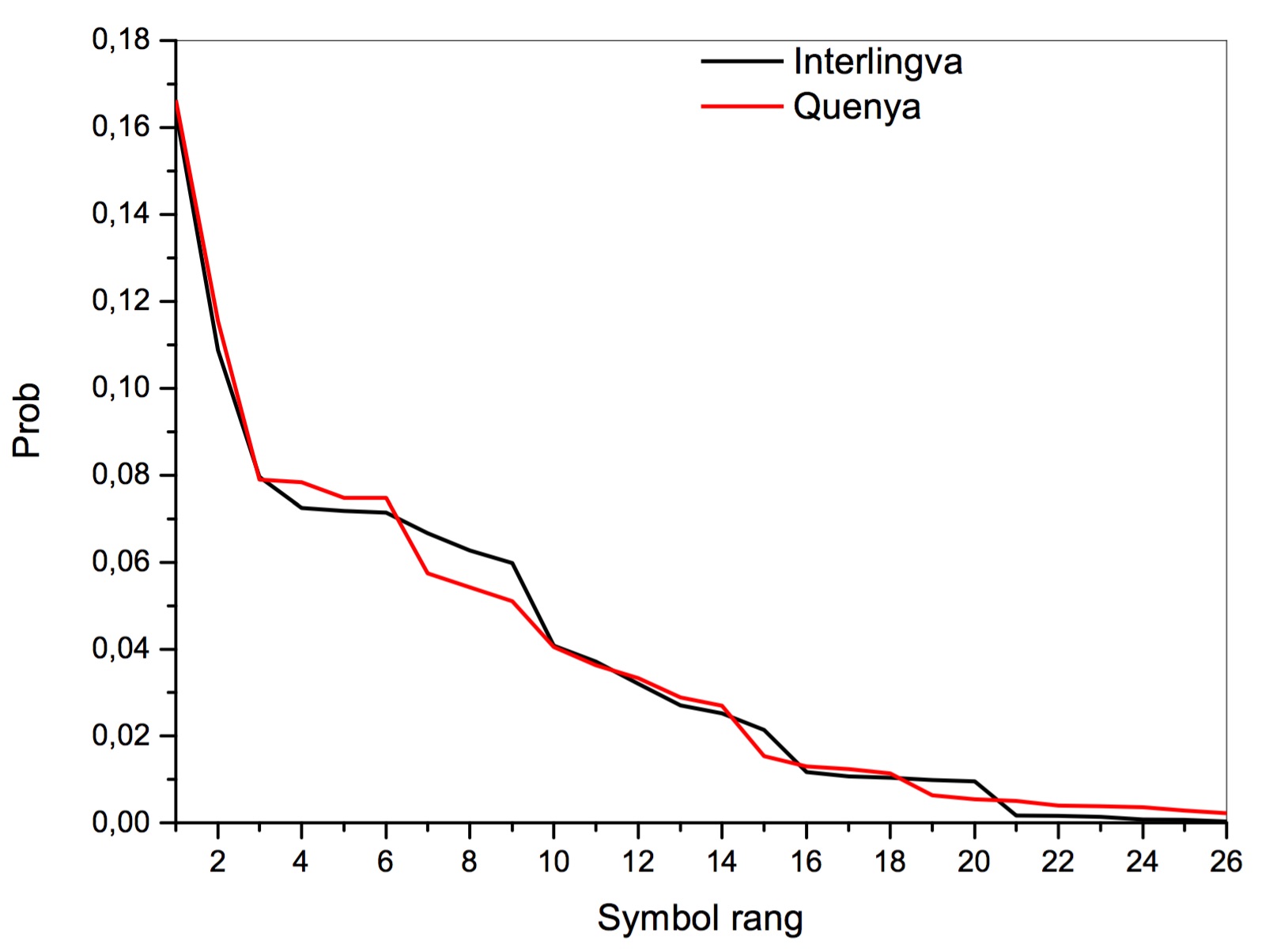}
\hfill
\caption{{\footnotesize Distribution of symbols of ordering some constructed languages}}
\label{fig_9}
\hfill
\includegraphics[width=70mm]{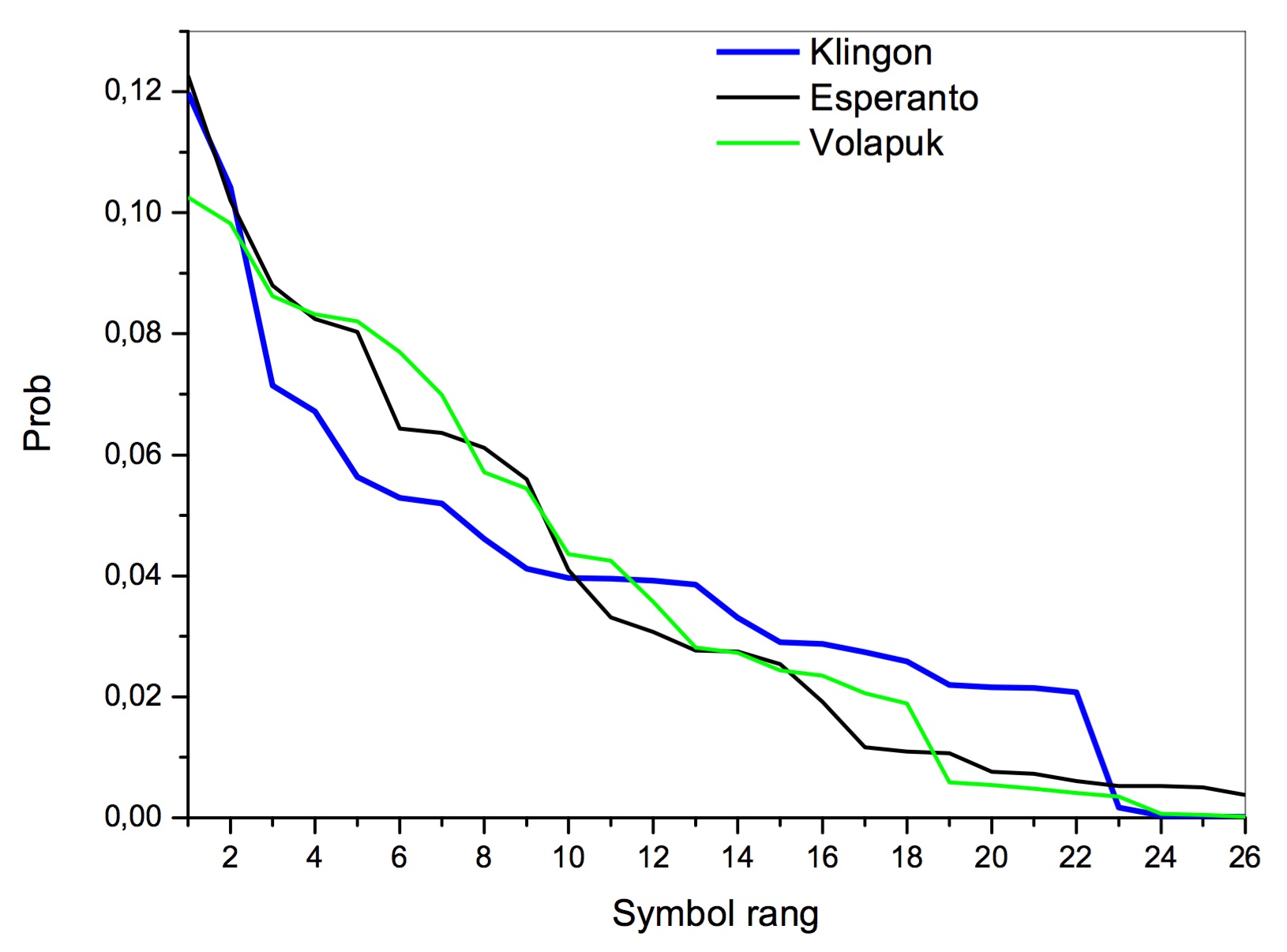}
\hfill
\caption{{\footnotesize Distribution of symbols of ordering some constructed languages}}
\label{fig_10}
\end{multicols}
\end{figure}

\begin{figure}
\begin{multicols}{2}
\hfill
\includegraphics[width=70mm]{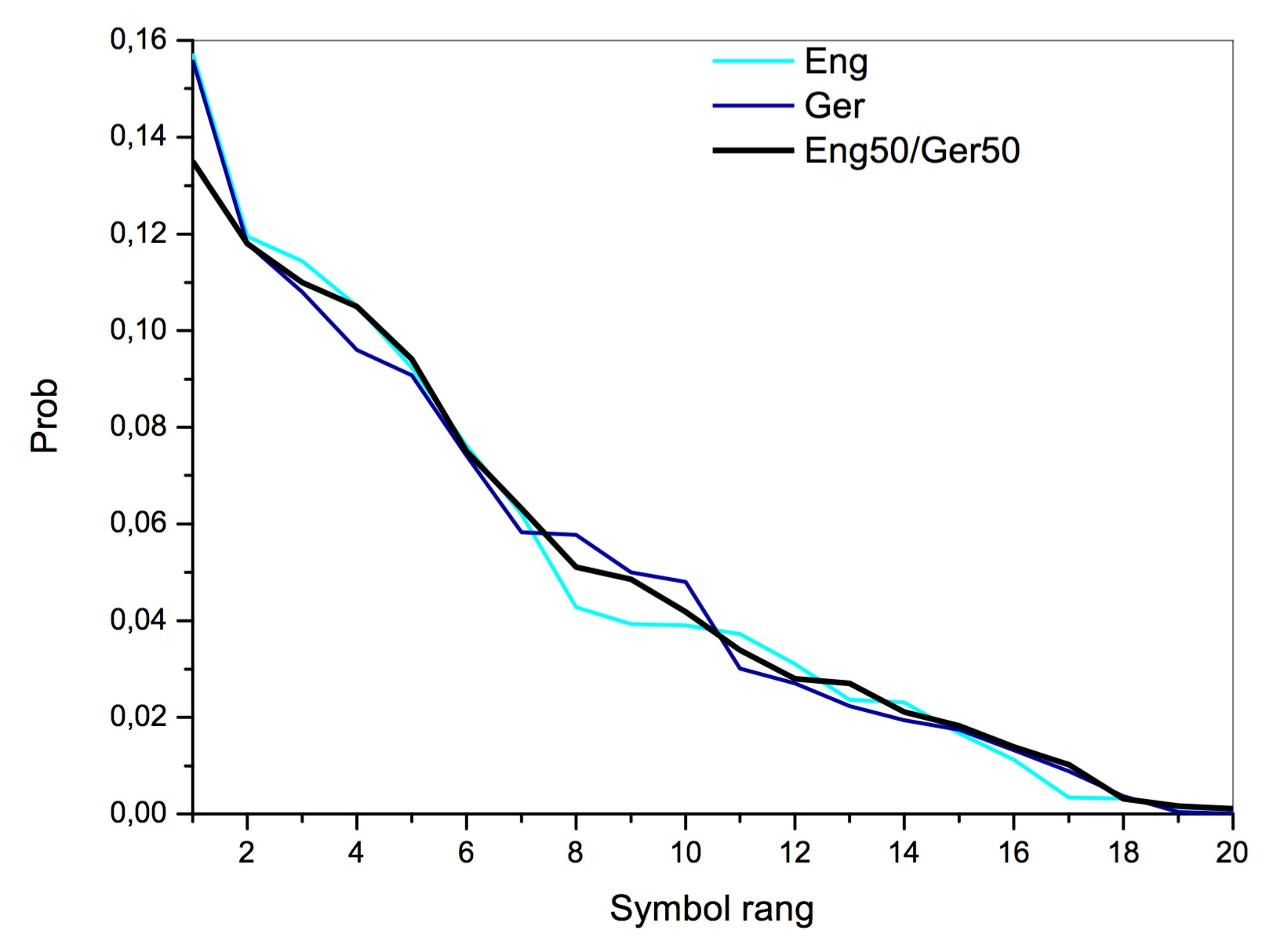}
\hfill
\caption{{\footnotesize The mixture of English/German texts}}
\label{fig_14}
\hfill
\includegraphics[width=70mm]{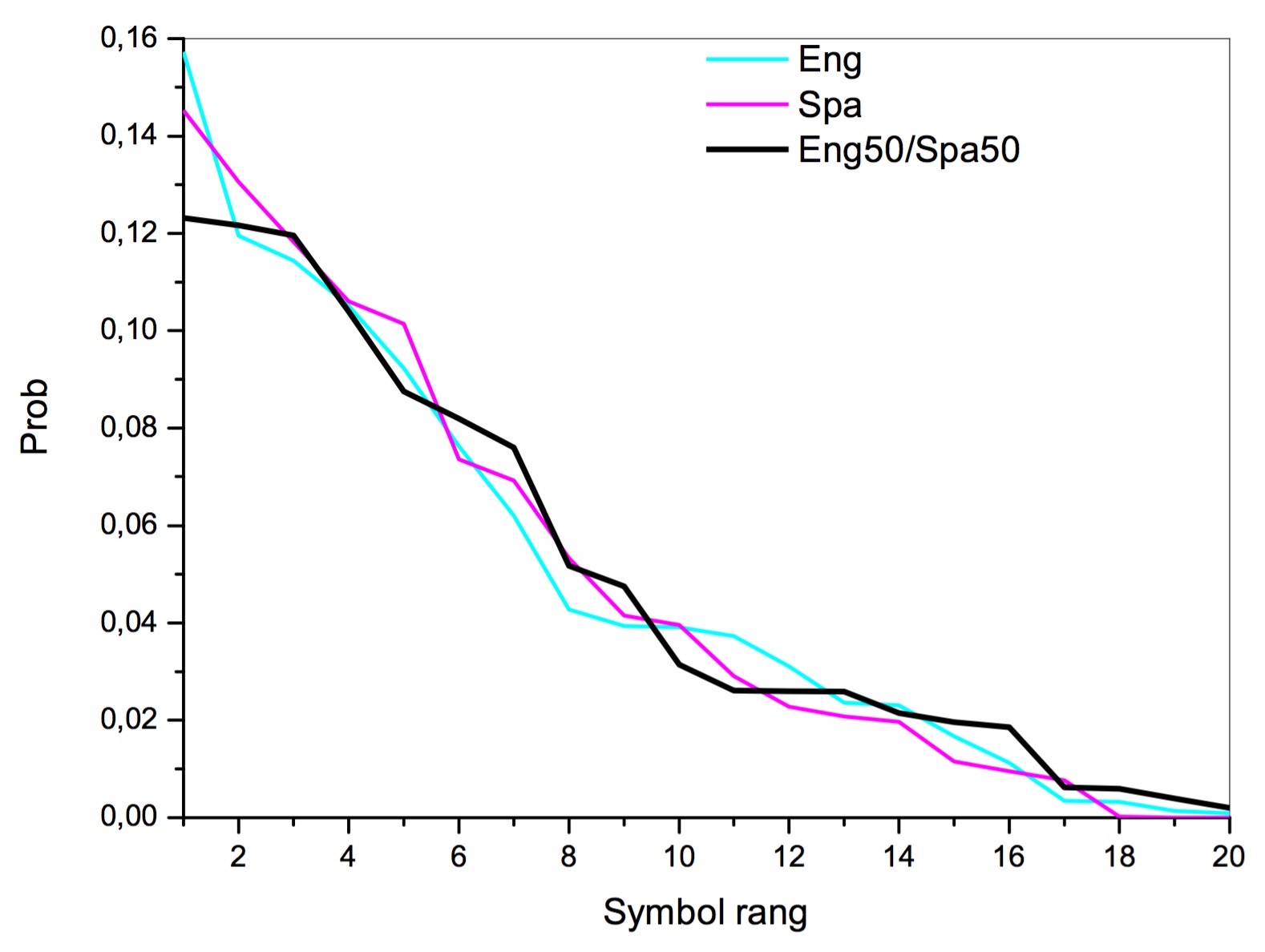}
\hfill
\caption{The mixture of English/Spanish texts}
\label{fig_15}
\end{multicols}
\end{figure}

\par However, the Takahashi transcription is even closer to the mixture of Latin and Danish languages in ratio 2:1, the distance between them is 0.9 (Fig.16)

\begin{figure}
\begin{center}
\begin{minipage}[H]{0.70\linewidth}
\includegraphics[width=1\linewidth]{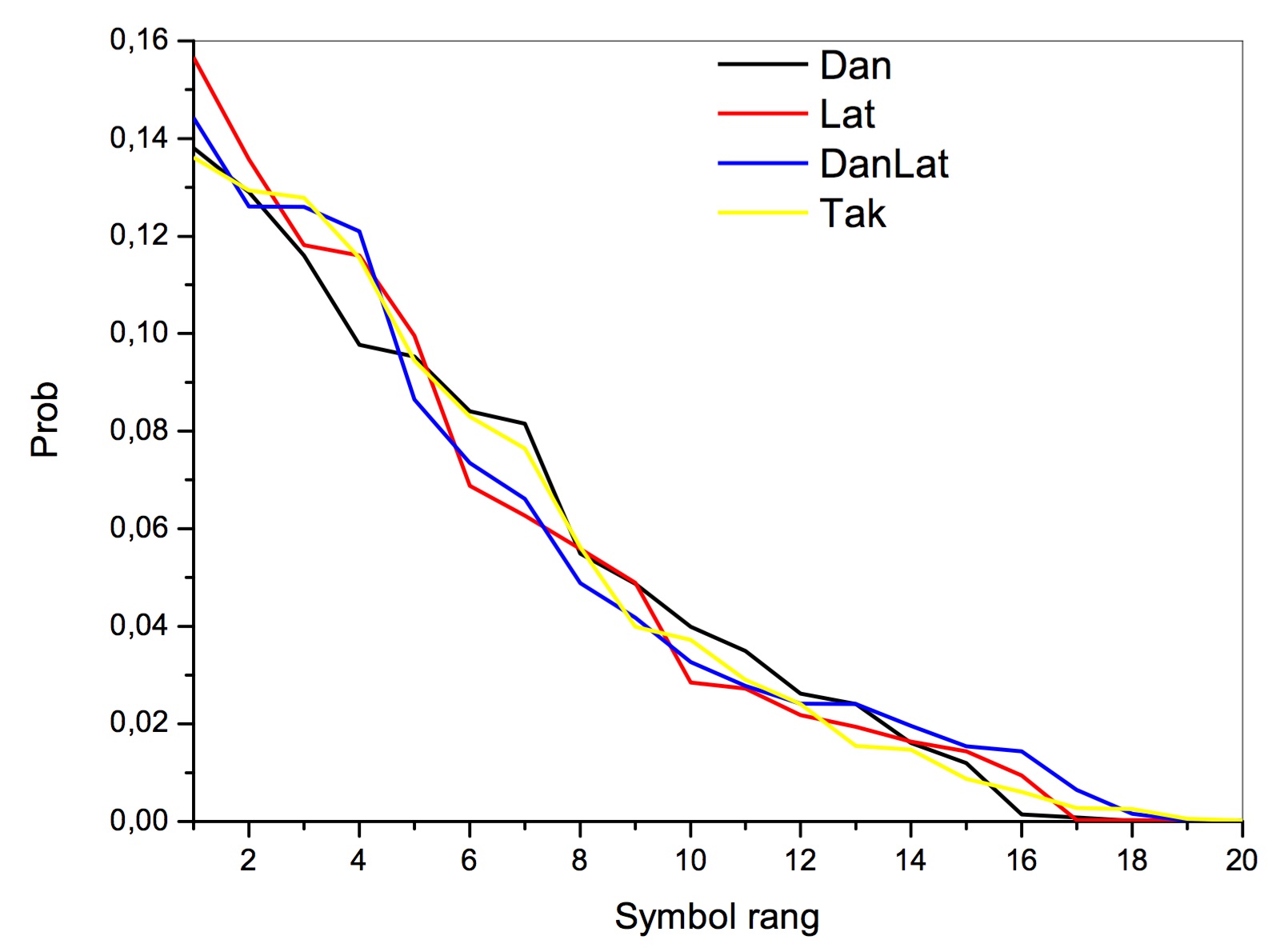}
\vspace{-4mm}
\caption{{\footnotesize The mixture of Danish-Latin texts}}
\label{fig_16}
\end{minipage}
\end{center}
\end{figure}
\par The analysis revealed that the texts written in different languages of the same group do not change its ordered frequencies symbol distribution when mixing texts in all possible proportions. For texts in different language groups either the mixture distribution changes, compared to the original distributions; or distances between distributions become more significant than the observed clustering level. We emphasize that these findings relate only to book-style texts (written by professional writers) where vowels and softening symbols had been removed. We also notice that while mixing English and Hungarian texts without vowels, the distance between pure and mixed texts remained unchanged (equal to 0.16), whereas the logarithmic model determination for the mixture was higher (0.99) than for the text components separately (Hungarian 0.96 and English 0.98, Fig. 17). This example shows that the preservation of determination of the languages mixture is not a necessary and sufficient condition for their affinity.
\begin{figure}
\begin{center}
\begin{minipage}{0.70\linewidth}
\includegraphics[width=1\linewidth]{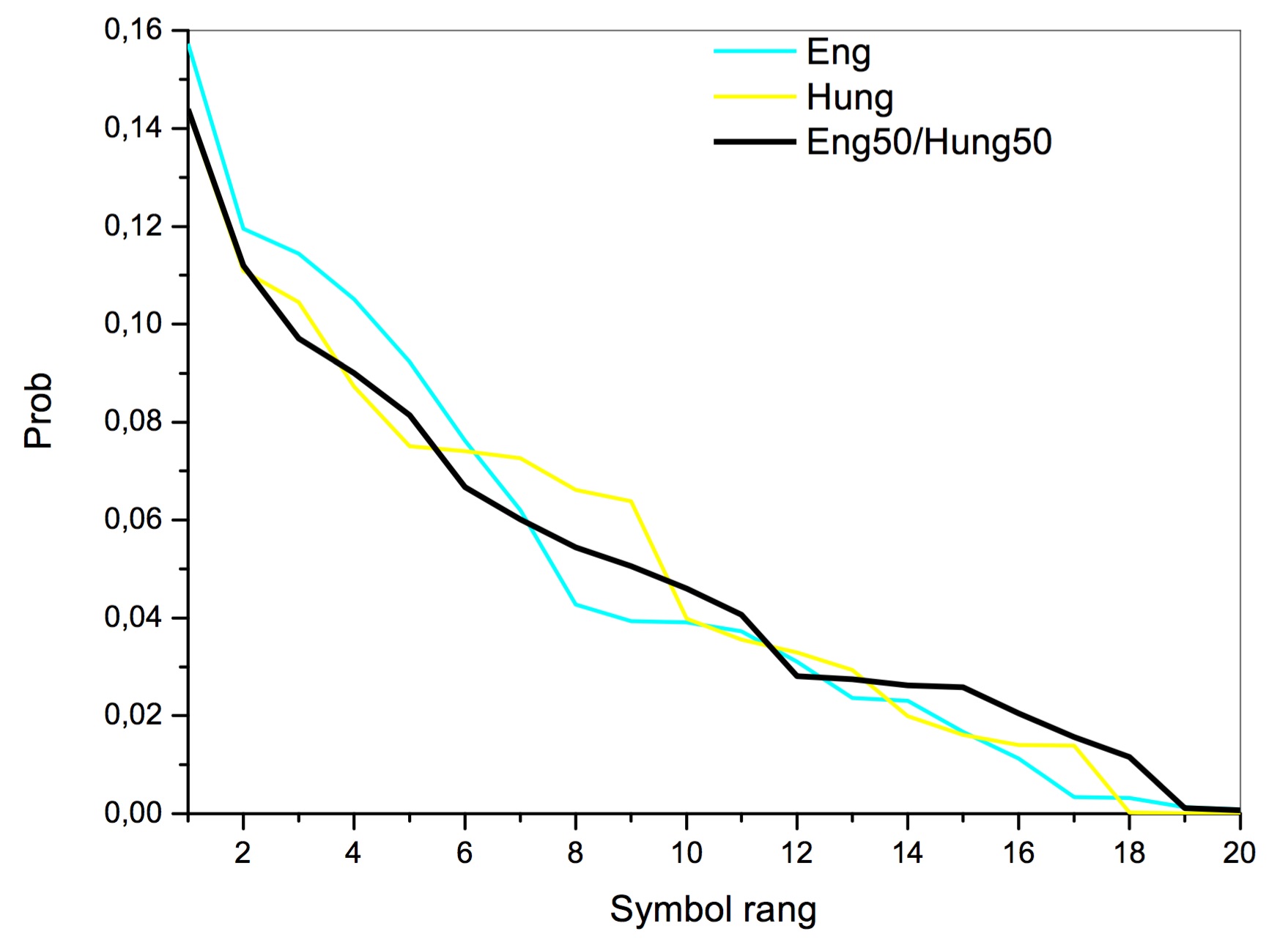}
\vspace{-4mm}
\caption{{\footnotesize The mixture of English and Hungarian texts}}
\label{fig_17}
\end{minipage}
\end{center}
\end{figure}
In \cite{ORIG} and \cite{OrOs} it is shown that the full distribution of symbols depends on professional texts specifics. Nuances between different texts are unimportant, determination level fluctuates at around 0.96 and difference between actual distributions and model dependency (\ref{1}) changes with the range of 0.09--0.13, that approximately coincides with similar characteristics of literary texts.
\par Consequently, we can assume that the text subject has no significant effect on the distribution of its ordered consonants and, therefore, conclusions drawn from the research on literary texts can be applied for texts on specialized topics.
\section{Identification of the Manuscript fragment language}
\par Distributions, constructed in Sec. 2 can be used to answer the question, where in the Manuscript text one language is mainly used and where there is a mixture. For this we need to apply a method proposed in \cite{VO} of identifying selective distribution functions for small samples. This method is described as follows. Suppose we have reference distribution functions (patterns) $F_{i}(x)$ and a fragment of the time series with selective distribution function $G(x)$.Then this fragment is considered to be a sample from the distribution $F_{j}(x)$ with the number

\begin{equation}\label{4}
j= \arg\min \|F_i(x)-G(x)\|
\end{equation}

\par The norm, in which the distance between features are calculated in is chosen with the aim of identification the minimum error in the test data set. It was found out in that for small samples with lengths about 50--200  values and for distribution densities, integrally deviating from each other in the $L_1$ norm at about 0.1--0.2, the best norm between the distribution functions is $L_1$: 

\begin{equation}\label{5}
\|F(x)-G(x)\|=\int |F(x)-G(x)| dx
\end{equation}

\par In order to facilitate comparison with previously calculated distances between the distributions, we use the $L_1$ norm between the probability densities. In other words, we calculate the distance between the references and the sample density distribution function $g_{n}(k,t)$, which is found from the sample of length $n$ by moving a frame with a single time step:

\begin{equation}\label{6}
p_{i}(n,t)= \sum\limits_{k=1}^{20} |f_{i}(k)-g_{n}(k,t)|
\end{equation}

\par The argument $t$ corresponds to a character number in the Manuscript text, with which the sample ends, and the index $i$ means a reference number in accordance with the language. There are two reference examples of the density distribution for the VM: the first is the empirical distribution $f_{i}(k)$  of one European language text using the Latin alphabet, the second example is two mixtures mentioned above (English/Spanish and Latin/Danish). Comparing simultaneous distances $\rho_{i}(n,t)$, corresponding fragments may be identified as being written mainly in the language with $\rho_{i}(n,t)=\min$.
\par However, one should keep in mind that this identification method is effective only when we have a complete set of references. Otherwise, the identification will be incorrect. Proposed method of identification is sufficiently accurate if there is a correct reference among the reference distributions. If it is not the case, the most similar reference will be found, but there is no guarantee of the correct recognition of course. Once again we would like to emphasize that we are talking about the European language that is the closest to the transcription Takahashi, instead of discussing which language the VM is actually written on. We next perform language identification by gradually reducing the length of the text fragment. 
\par The whole text in Takahashi transcription is closest to the mixture of Latin and Danish. The distance to the mixture reference equals to 0.09, which is the lowest value among all possible combinations of modern European language pairs, including Latin.
\par We further divide the VM into four parts approximately with 45 thousand symbols each. It turns out that the first two parts are ultimately close to the Danish language reference with a distance of 0.08, the third part is close to Latin with distance 0.10, and the fourth one is close again to a mixture of Latin and Danish with distance 0.07. It should be noted that this result does not mean that the first half of the text is written only in Danish. The fact is that, among the considered references, Danish turned out to be even closer than the reference ``70$\%$ Latin and 30$\%$ Danish", which was used as a mixture reference. Just subsequent text fragments specification allows to discover more accurately, which language each fragment is mainly written in.
\par By further reducing the length of the considered fragment the distance to the closest reference increases with the growth of the statistical uncertainty  of the sampling distribution and for the fragments of the length 10000 symbols this measure equals to 0.1. Notice that in all cases the distance to the closest reference was smaller than 0.13 which corresponds to the discovered distance of the language groups splitting. 9 out of 17 fragments were identified as written in Dutch (these were the fragments 1,2,3,4,5,7,8,12 and 15 with the length of 10000 symbols), 6 -- in Latin (fragments 6,9,10,11,14 and 16), 2 fragments (13 and 17) were identified as written in Dutch/Latin mix. In addition to that, the first position in 10 of the cases was taken by the letter ``O", symbol ``E" was encountered 4 times and ``T" - 3 times. The second place took the symbols ``T" (7 times), ``E" (6 times), ``O" (3 times), ``W" (once). It shows that the distribution of the ordered frequencies can not provide the precise information about the most common symbol, which prevents from decoding of the text under the assumption of its exact language.
\par Consider now the fragments of the length 1000 digits (approximately 2.5 pages of the script). The same algorithm as above mentioned allows us to argue which European language each fragment is close to, regarding the ordering of its symbols. The corresponding language ``coloring" is shown in the Fig 18. However, it should be taken into consideration, that if the distance to the nearest reference distribution becomes too large, it is highly likely that the intended distribution is missing in the library. It seems interesting that along with the expected Dutch and Latin one can observe German and Spanish. As it was already mentioned, the distributions of the ordered frequencies correspond to the language group, therefore it makes sense to identify Spanish and Latin as a single Roman language group, and German and Dutch as a single German group. However we should mention, that about 15 $\%$ of the fragments were identified at the low level of confidence, because the distances to the closest references turned out to be sufficient (more that 0.15). It can be the case, that the reference distributions were constructed with the help of contemporary texts; another reason can be that we were not able to find the actual reference.
\begin{figure}[ht]
\begin{center}
\begin{minipage}[ht]{0.70\linewidth}
\includegraphics[width=1\linewidth]{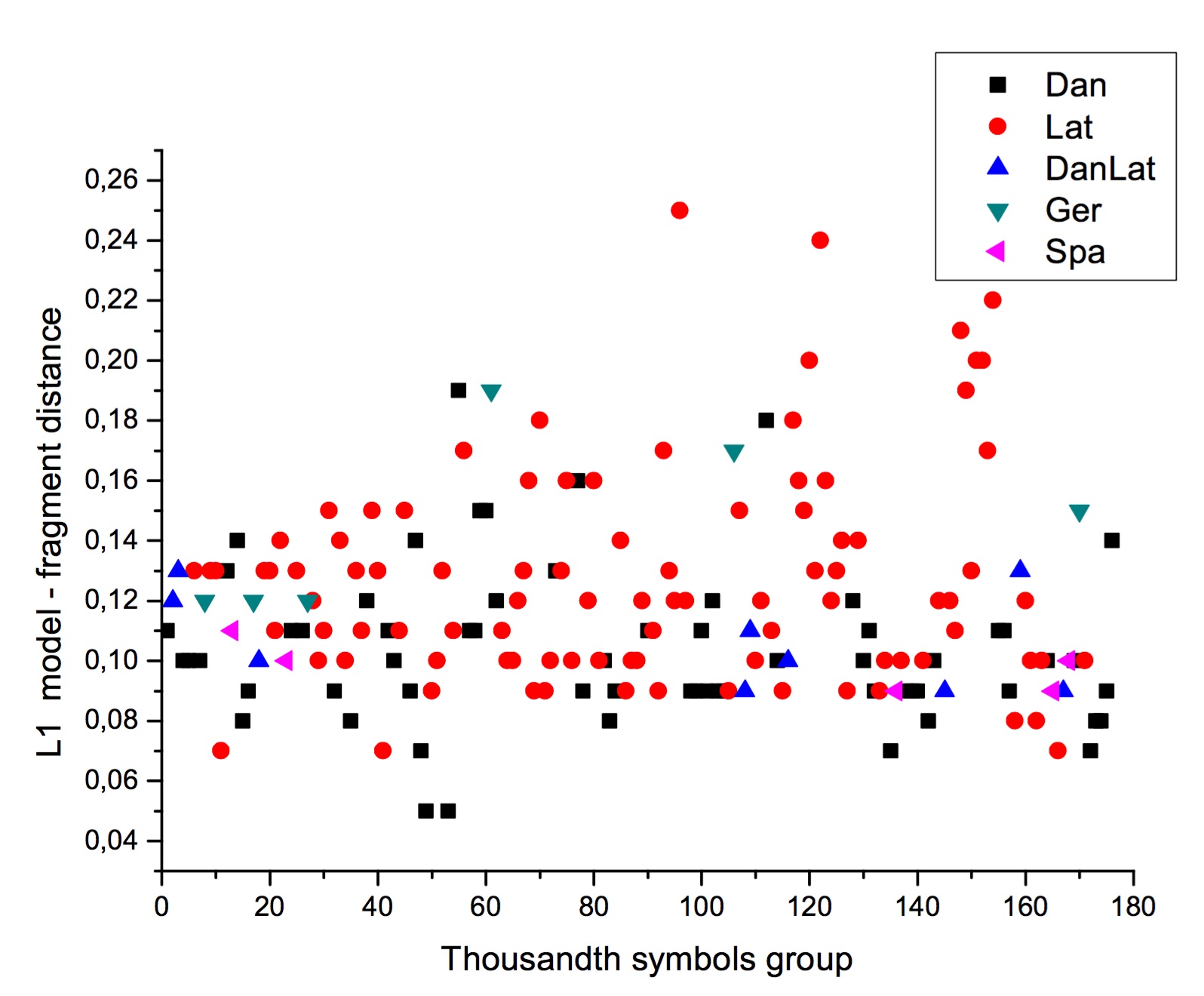}
\vspace{-4mm}
\caption{{\footnotesize The closest distributions to the fragments of the length 1000 digits}}
\label{fig_18}
\end{minipage}
\end{center}
\end{figure}

Nevertheless, an abundance of arguments support the fact that the text to be written in two or more European languages.
\section{Spectral portrait of Voynich Manuscript}
Considering matrix $P_{ij}$ of empirical conditional probabilities of the fact that at some point of the text is a symbol $j$, provided that to the left of it is a symbol $i$ This matrix is described by double-letter distribution $F(ij)$ and single-letter distribution $f(i)$ as follows: 

\begin{equation}\label{7}
P_{ij} = \frac{F(i,j)}{f(i)}, f(i)=\sum_{j} F(i,j)
\end{equation}

\par According to (\ref{7}) it follows that matrix $P_{ij}$ has eigenvalue, which is equal to unit, and corresponding eigenvector is a symbol distribution $f(i)$. Other eigenvalues of this matrix characterize the stability of the frequencies of pairs of letters for fragments of the text. Conforming to S.K. Godunov \cite{Godun}, the value $\lambda$ belongs to the $\epsilon$-range $\Lambda_\epsilon(P)$ of matrix $P$, if consists such disturbing matrix $\Delta$ , that $\bigl\|\delta \bigr\|\le \epsilon\bigl\|P\bigr\|$ and $\det(\lambda I-p-\Delta)=0.$
\par The resolvent of the matrix $P$ is defined as 

\begin{equation}\label{8}
R(\lambda)=(\lambda I-P)^{-1}
\end{equation}

\par Using the concept of resolvent the Epsilon-range is determinated this way: value $\lambda$ appended to the Epsilon-range $\Lambda_\epsilon(P)$ if the following condition is valid:   

\begin{equation}\label{9}
\bigl\|R(\lambda)\bigr\| \ge \frac{1}{\epsilon \bigl\|P\bigr\|}
\end{equation}

\par In practice, one of numerical algorithms using to determine regions in the complex plane of parameter $\lambda$  is based on formula (\ref{9}). The closed smooth curves $\gamma_\epsilon$ that are isolines of the $\epsilon$-spectrum are interesting in the research of the spectrum of point locations. The circuit $\gamma_\epsilon$ separates the whole $\epsilon$-spectrum $\Lambda_\epsilon (P)$ into two parts lying inside and outside. The option $K_\gamma (P)$ of the dichotomy is evaluated by the norm square of the resolvent (\ref{9}) at the given curve:

\begin{equation}\label{10}
K_\gamma (P) = \frac{\bigl\|P\bigr\|^2}{l_\gamma} \oint_{\gamma} \bigl\| R(\lambda)\bigr\|^2 d\lambda.
\end{equation}

\par Here $l_{\gamma}$ is the length of the contour $\gamma$ . The value $K_\gamma (P)$ is selected as the accuracy indicator of the separation of the spectrum. If certain curve $\gamma$ has no points of the spectrum $\lambda (P)$, the norm of the resolvent of a on such a curve is finite
$\bigl\|R(\lambda)\bigr\|_{\gamma} < \infty $ along with the integral over this curve.
\par If some eigenvalues are belonged to the region bounded by the curve $\gamma_\epsilon $  it is natural to consider them coincident with the specified accuracy $\epsilon$ . In this case the subspace with basis consisting of eigenvector and adjoined vector for such multiple eigenvalue is an invariant subspace for the operator $P$.
\par It is convenient to consider the radial dichotomy that is dichotomy is shaped by the curve $\lambda = re^{i\phi}$ with a fixed value $r$.
In this case the option $K_r(P)$ of the dichotomy is the norm of the Hermitian matrix $H_r(P)$ with the integral representation

\begin{equation}\label{11}
H_r(P)=\frac{1}{2\pi}\int\limits_0^{2\pi} (P^+ - re^{-i\phi}I)^{-1}(P-re^{i\phi}I)^{-1} d\phi, \quad K_r(P)=\bigl\|P\bigr\|^2 \bigl\|H_r(P)\bigr\|
\end{equation}

\par The integral (\ref{11}) converges only if on the circle  $\lambda = re^{i\phi}$ there are no eigenvalues of the matrix $P$ . This formula is used in finding the numerical $\epsilon$-spectrum of the matrix in the form of level lines for the $L_2$-norm of the resolvent. They are given below. 
\par It makes sense to compare the spectral portraits of the matrices (\ref{7}) for two transcriptions of the VM, as well as for texts of the Germanic and Romanic groups without a vowel. The calculation results are shown on the figures 19--22. The areas with the same color have eigenvalues of the matrices if the elements of these matrices are known with the precision noted in the legend.
\par All matrices of the form (\ref{7}) have one separate eigenvalue equal to one. The remaining eigenvalues form a structure which are characteristic of one or the other languages. It makes sense to consider real eigenvalues, kernel close to zero and large in absolute value complex eigenvalues. For all European languages the area of the spectrum is approximately limited by the circle with the radius 0.2 (the green region is shown on the Fig.19 and Fig.20). According to the paper \cite{ORIG} the area of the spectrum for the texts with the full alphabet has the form not of a circle but of an ellipse with semimajor axis equal to approximately 0.5 and semiminor axis still equal to 0.2. 
\begin{figure}[ht]
\begin{center}
\begin{minipage}[ht]{0.70\linewidth}
\includegraphics[width=1\linewidth]{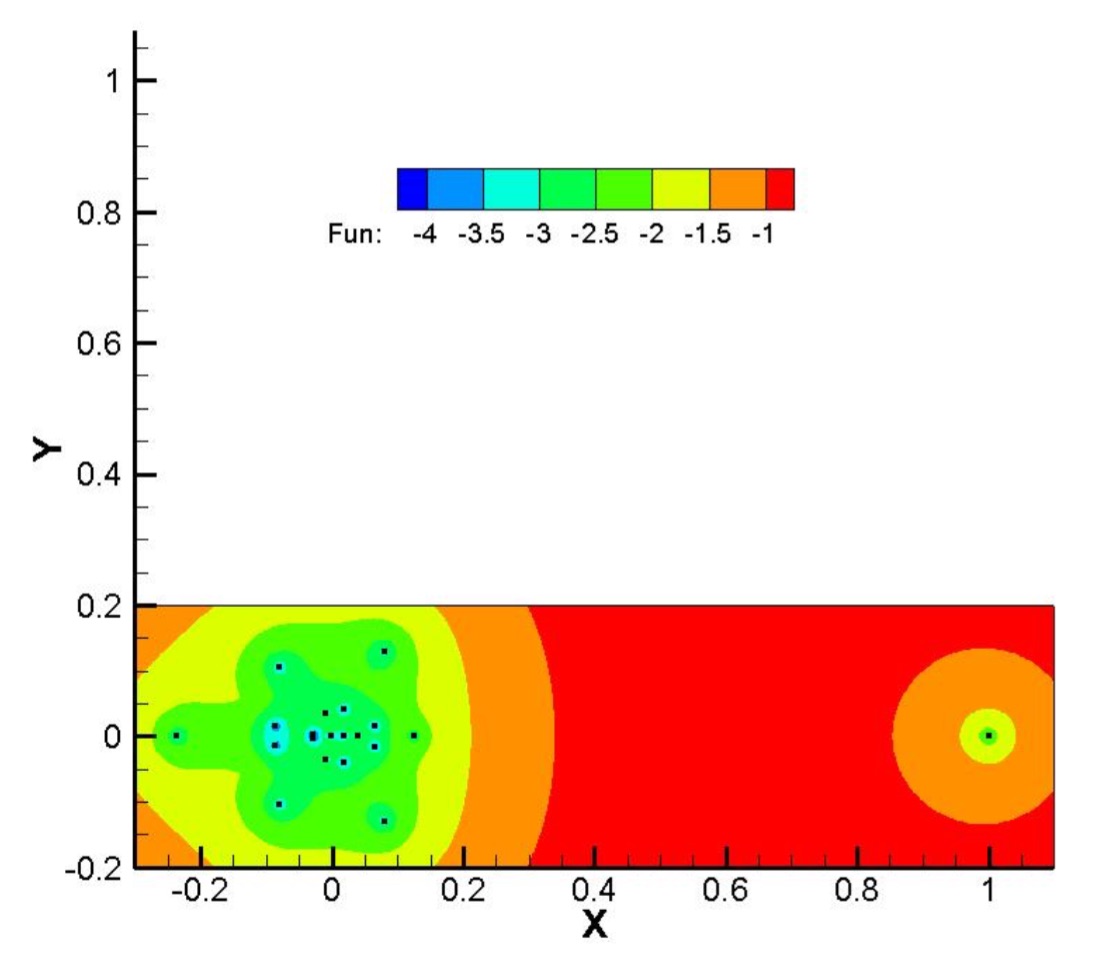}
\vspace{-4mm}
\caption{{\footnotesize Spectral portrait of text without vowel is written in English }}
\label{fig_19}
\end{minipage}
\end{center}
\end{figure}

\begin{figure}[ht]
\begin{center}
\begin{minipage}[ht]{0.70\linewidth}
\includegraphics[width=1\linewidth]{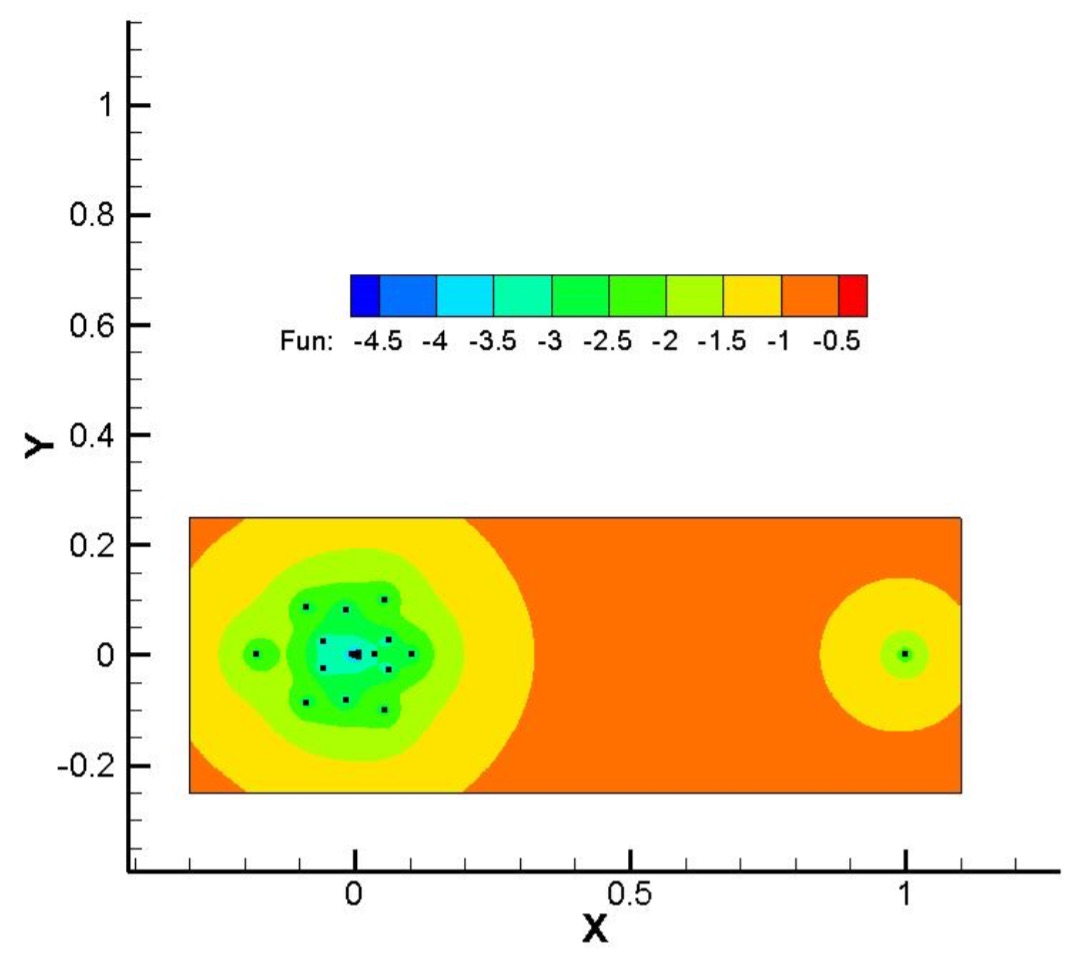}
\vspace{-4mm}
\caption{{\footnotesize Spectral portrait of text without vowel is written in Latin }}
\label{fig_20}
\end{minipage}
\end{center}
\end{figure}

\begin{figure}[ht]
\begin{center}
\begin{minipage}[ht]{0.70\linewidth}
\includegraphics[width=1\linewidth]{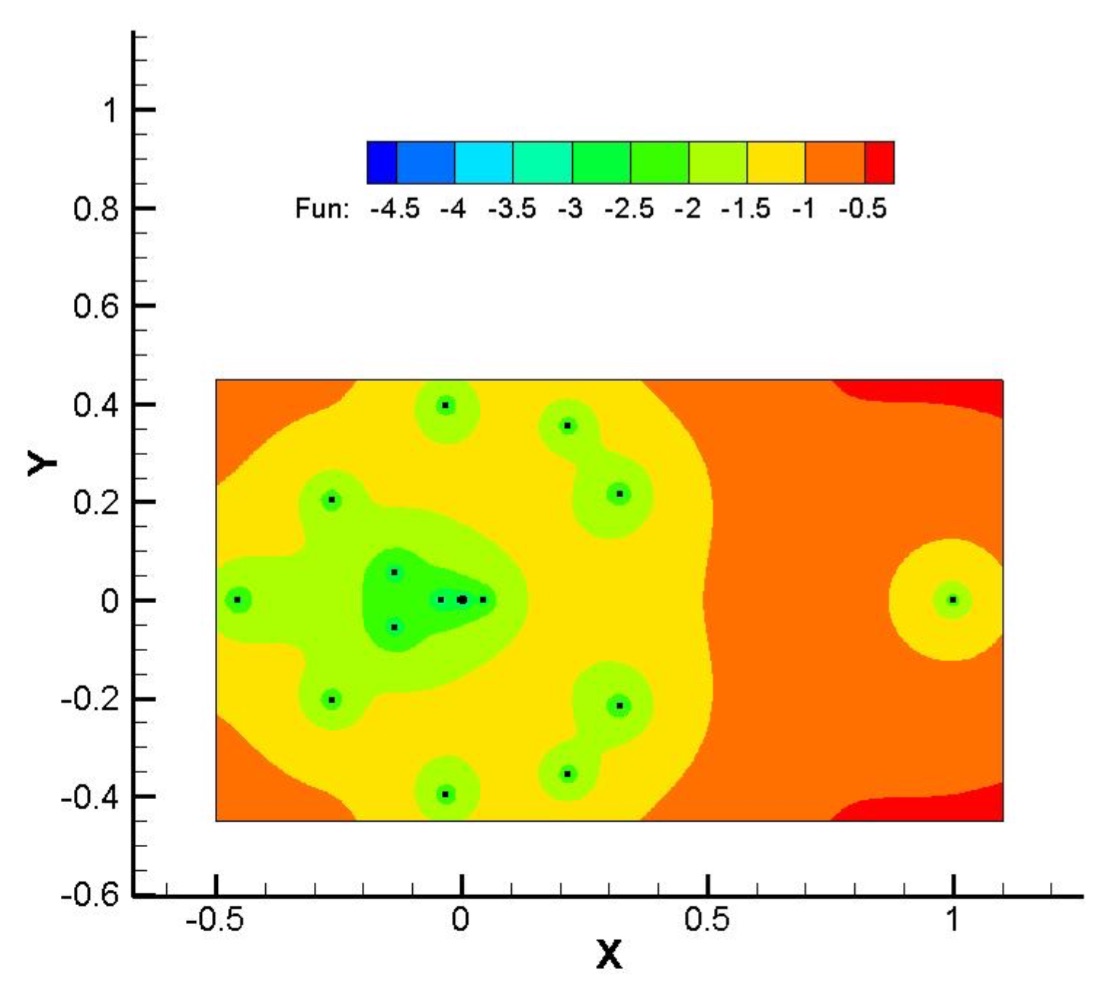}
\vspace{-4mm}
\caption{{\footnotesize Spectral portrait of the transcription EVA }}
\label{fig_21}
\end{minipage}
\end{center}
\end{figure}

\begin{figure}[ht]
\begin{center}
\begin{minipage}[ht]{0.70\linewidth}
\includegraphics[width=1\linewidth]{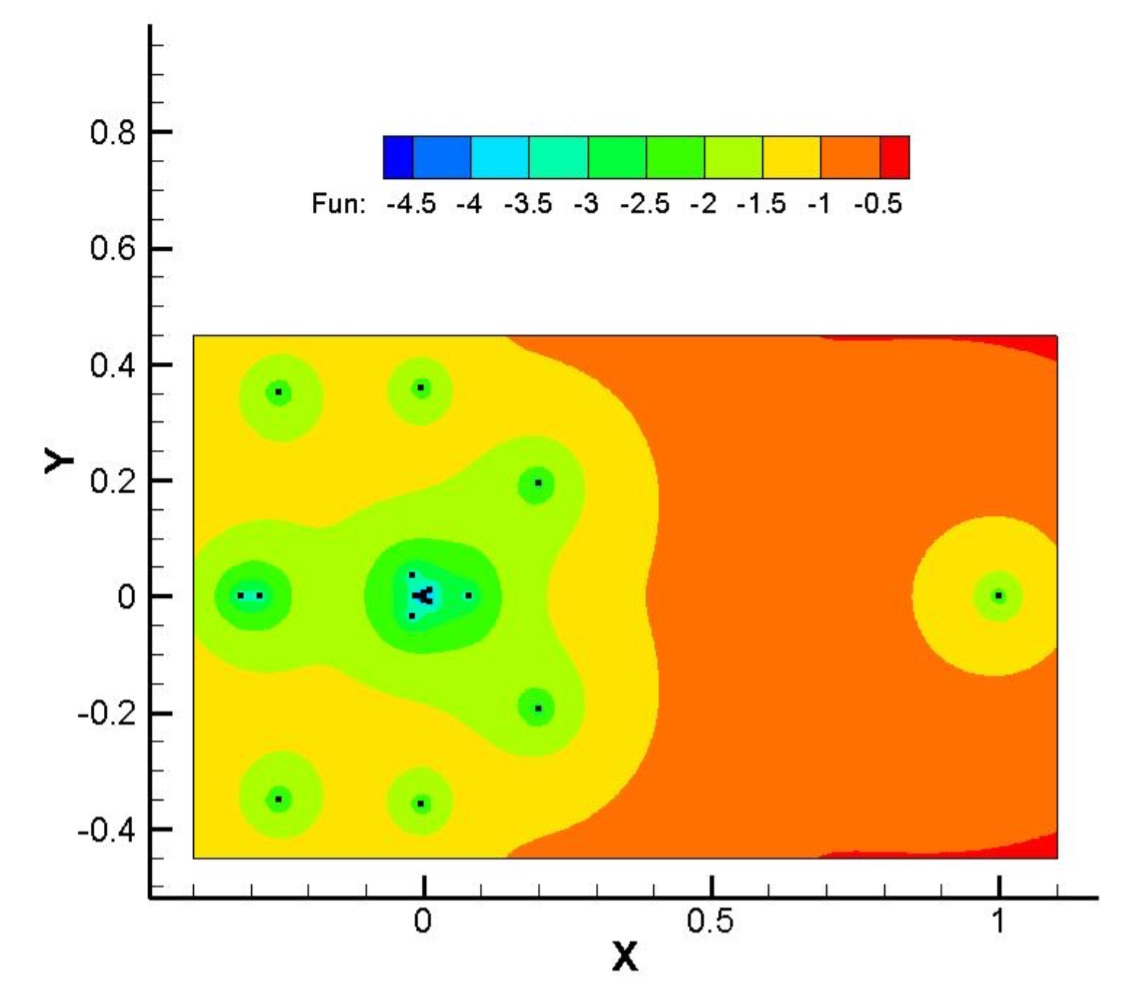}
\vspace{-4mm}
\caption{{\footnotesize Spectral portrait of the transcription Takahashi }}
\label{fig_22}
\end{minipage}
\end{center}
\end{figure}

\par Comparing figures 19--20 and figures 21--22, we see that the regions with equal accuracy are markedly different in finding eigenvalues of the matrices (\ref{7}) for the VM and conventional texts (it conserns both texts in the full alphabet and without the vowel). For the both transcriptions of the VM the circle (it is not an ellipse!) of the location of the eigenvalues has approximately two times larger radius than for natural languages. It has fundamental importance. The spectrum of the EVA shifted to the left and spectrum of the Takahashi to the right. The difference of the spectral portraits of the transcriptions corresponds to differences in the distributions of ordered frequencies (the red and the green curves are shown at the Fig.1). The bulge of these curves vary in opposite phase. Notably, the both transcriptions have five disjoint spectral zones of equal accuracy $10^{-2}$ (the light green is shown at the figures 21--22).
\par
It is important to emphasize that all these arguments are fundamentally different, i.e. they express the peculiar properties of independent statistics, indicating that the interpretation of the VM as part of the composite manuscript is acceptable.
\section {Voynich Manuscript structure} 
\par At the end of our research let us speak about the fundamental possibility of statistical detection of the manuscript fragments written in ``a similar manner", if such a term could be applied to the text in an unknown language. Back to the Introduction it is not guaranteed that the Voincih Manuscript has the correct ordering. The statistical research of the fragments of the Voincih Manuscript from the Chapter 5 in a sliding window found that the Voincih Manuscript cannot be treated as a single document. It can be treated as a number of independent works written for example by the ``brotherhood" with their secret language. In this section we look at the statistical aspects of the VM analysis -- the homogeneity of individual sheets. Following the technique \cite{OrOs} we will use the functional distance between the distributions of characters in their alphabetical ordering to solve this problem. We analyze the transcription Takahashi to avoid uncertainty.
\par The analysis of literary texts in European languages, which was carried out in work \cite{OrOs}, showed that between two fairly large (10 thousand characters) texts by the same author on the same language the distance in the sense of the norm $L_1$ for alphabetical ordered distributions of characters is 0.03--0.07. This distance for different authors is in the range of 0.04--0.13, and for different languages regardless authors the distance between the texts is 0.20--0.50. With regard to the VM \cite{Shailor}, it makes sense to check how close the distributions of conditioned parts of the manuscript in accordance with the existing illustrations. Without pretending to be original, we separate the traditional ``botanical" part of the VM (sheets 1--57 of the Manuscript), ``female body" (sheets 75--85) and ``astrology" (sheets 103--116). On the possible violations of the sequential numbering of the sheets indicates that the sheets 87, 90, 93--96 have clearly ``botanical" type, the sheets 58, 68, 69 are similar to the subsequent ``astrological" part, and the sheets 65, 66 would be convenient to refer to the ``women's bodies". The thematic belonging of the sheets 67, 70--73, 86, 88, 89, 99--102 ornamented by some ``mortars" and plants. Table number 4 shows the distance between the separated parts of the VM. It understood as the distance in the norm $L_1$ between the distributions of the ordered frequencies.

\begin{table}
\caption{{\footnotesize The distances between the distributions of characters in the VM}}
\begin{center}
\begin{tabular}{|g|g|g|g|g|}
\hline
& {\bf Botany} & {\bf Bodies} & {\bf Astrology} & {\bf Mortars} \\
\hline
{\bf Botany} & \cellcolor{darkishgreen} & 0.30 & 0.20 & 0.25 \\
\hline
{\bf Bodies} & & \cellcolor{darkishgreen} & 0.18 & 0.27\\
\hline
{\bf Astrology} & & &\cellcolor{darkishgreen} & 0.20\\
\hline
{\bf Mortars} & & & &\cellcolor{darkishgreen} \\
\hline
\end{tabular}
\end{center}
\end{table}

\par The distances from the Table 4 are characterized by texts written in different languages. This texts may be the same or different. For example, a novel written in English and this is novel written in French. The most important that the text written in the same language differ not more than 0.13 in the norm $L_1$. The length of each of the fragments of the VM more than 10 thousand signs, so that the conclusion of the linguistic disparity of the VM seems reasonable. Note that half of each part is different from the other half on 0.10, so the allocation of parts of the VM in accordance with drawings is logically.
\par Let's analyze, which parts of the VM are close to the individual sheets, what do not have an unambiguous interpretation. One sheet of the VM has two pages and contains from 500 to 2000 symbols, depending on size of the images. The sheet of ``its" part is separated from this part by a distance of 0.10--0.30, and by ``foreign" part to 0.25--0.50. Let's try to identify the affiliation of the above ambiguous sheets by the proximity of their distribution to parts of the VM.
\par The ``Botanical" part is the nearest to the following sheets: 69 (the distance to the standard is 0.36), 86 (the distance is 0.16), 87 (the distance is 0.30), 93--96 (the distance is 0.15).
\par To the ``Bodies" are close sheets 65 (the distance is 0.32) and 66 (the distance is 0.17).
\par To the ``Astrology" part are close the sheet 58 (the distance is 0.33).
\par The rest ambiguous sheets 67, 68, 70--73, 88--90 are close to the ``Mortar" part (99--102 sheets) with distances 0.17--0.32.
Thus, the 4 sheets of the 23 have been identified not as it should be in accordance with drawings. This lists are 68 (mortars instead of the expected astrology), 69 (botany instead of astrology), 86 (botany instead of mortars) and 90 (botany instead of mortar). But this fact is not so important as the fact that some 4 sheets (not all of them are those that are identified as ``wrong") are spaced from closest standards at very large distances exceeding 0.32 (the last distance decile to recognize ``significantly" sheets). This indicates insufficient reliability about their output.  The share of such fragments among considered 23 sheets is 0.16. The accuracy of the similar identification of literary texts \cite{OrOs} is 0.15, which is close to estimates made for the VM and typical for this method. So, the marked fragments may claim to a thematic association with the fragments with the distance significantly less than to others. It is essential that these distances are approximately equal distances between the parts of fragments that are considered internally homogeneous. This demonstrates the correctness of the proposed merger. Thus, the new point of view on the VM as a manuscript written in several languages, not only, but as no one, but two or three different manuscripts is possible.
\par It is necessary to indicate the accuracy of the results presented in this paper. We work on statistical pattern recognition by comparison with the standard. The critical point is the accuracy with standard itself known. In this case standard refers to the probability distribution of text characters. If the text is made up of $N$ signs and written by alphabet of $n$ signs, the distribution of these characters in the text is determined with a precision $\epsilon$ that find numerically from equation \cite{Orl}:

\begin{equation}\label{12}
\frac{u_{1-{\epsilon}/2}}{\epsilon}= \frac{\sqrt{N}}{\Sigma_N(n)}, \quad \Sigma_N(n)=\sum_{j=1}^n\sqrt{f_N(j)(1-f_N(j))}
\end{equation}

\par There $u_{\gamma}$ is a $\gamma$-quantile of the normal distribution. The quantile of Student's distribution order $N$ at large values of $N$ is approximated by corresponding $u_{\gamma}$, and $f_{N}(j)$ is the empirical frequency of symbol $j$ in this text with length $N$. In particular, for the logarithmic ordering model (1) when $n=20$ and $o=0$ the value of the sum in (\ref{12}) is equal to 3.93; in relation to the VM with the number of signs $N=1.7\cdot 10^{5}$ its actual distribution leads to $\Sigma_{N}(n)=3.65$. The right side of the equation with respect to $\epsilon$ (\ref{12}) for a theoretical model is equal to 105, and for the VM is equal to 113. These values correspond to similar accuracies $\epsilon=0.02$, which differ in the third decimal place. It is similarly found out that the accuracy of the frequency distribution for a single page (1500 characters) is 0.1. Consequently, the differences between the distributions of the fragments at the level of 0.08--0.13, and sheets of fragments at the level of 0.20--0.40 are not caused by statistical noise of samples, they caused by objective reasons.
According to the estimates of accuracy, reliable field at the spectral portraits of the VM in the figures 21--22 correspond to the light green legend icons. Thus, the difference between the samples with the specified accuracy of statistical estimates is well defined.
\section{Conclusion}
The results of presented statistical investigations can be summarized as follows.
\par The classification of the Indo-European languages into distinct groups can be performed according to a formal statistical procedure, i.e., pairwise clusterization of symbol frequencies distributions in texts without vowel letters. Within these sub-groups languages can be mixed together without any significant change in the frequency distribution. It should be noted however that this rule is not universal, e.g. languages of the Uralic family do not clasterize well. It was shown that the distribution of Hurst exponents can be treated as a language invariant. Spectral portraits of texts written in the Indo-European languages have close similarities in layout of eigenvalues.
\par Concerning the Manuscript, it seems most plausible that it was written in two languages having the same alphabet without vowel letters: 30$\%$ of the text is written in one of the Germanic languages (Danish or German) and the rest 70$\%$ -- in one of the Romance languages (Latin or Spanish). The reasoning behind this statement relies on the observation that the distribution of symbol frequencies for the Manuscript resembles the features of meaningful texts, but the distribution of Hurst exponents for time series of distances between pairs of identical letters behave completely different compared to texts written in one language (natural or constructed). In addition, distances between alphabet distributions of large fragments of the Manuscript are also typical of texts written in several different languages.
\par Proposed statistical techniques can be used to refine the pagination of the Manuscript according to its thematic sectioning. However, it remains unclear whether the parts of the Manuscript are distinct works or the same work: large distances between the fragments are typical for the case of different languages, not different works; in the latter case these distances are much lower.
\par Yet the most intriguing questions ``what are the origins of the Manuscript, who wrote it, what is it about, and, most importantly, why it was written" cannot be answered here. Only a rigorous and accurate translation of the Manuscript might shed some light on these issues.
We can propose (as a historical reconstruction) only a conjecture about the original purpose of the Manuscript and the creation of its peculiar alphabet. Probably, the alphabet itself was designed by a small group of scholars (presumably alchemists) on the basis of contemporary script. After some practice they reached desired fluency in it (judging by the ease with which the Manuscript was scribed); several treaties were written. However, this is not the only possibility, e.g. they might have prepared a draft of translation from ordinary language into the cipher, and then made a copy of it, which is known now as the Manuscript.
\par After some time this scholar group vanished for some unknown reason. But some of their works (botanical, anatomical and astrological tractates) survived. They were stored, probably, together and were not used by other scholars, because no one was able to decipher them. All consequent owners of these treaties did not have a clear understanding of what was written there. Probably at this time several pages were accidentally placed in the wrong order, and to prevent further shuffle page numbers were added. The rest of the story is well-known: all texts were sent to the library of Collegio Romano where they were discovered by W. Voynich who gave the first detailed description of the Manuscript.
It is obvious that the results presented here tell nothing about the possible subjects of the Manuscript. We might never find out what exactly happened to the scholars who created it in an attempt to conceal their knowledge. But we hope that eventually the Manuscript itself shall be interpreted, with the help of statistical methods proposed here or any other research techniques and approaches.


\end{document}